%% file: hmi_industry40_survey.tex
\documentclass[acmsmall]{acmart}

\newif\ifsubm
\submfalse 

\usepackage[colorinlistoftodos, textwidth=2.0cm]{todonotes}

\RequirePackage[index=true]{acro}
\NewDocumentCommand\acrodef{mO{#1}mG{}}{\DeclareAcronym{#1}{short={#2}, long={#3}, #4}}

\include{acronyms}

\usepackage{array,ragged2e}
\usepackage{tabularx}
\newcolumntype{L}[1]{>{\RaggedRight\arraybackslash}p{#1}} % linksbündig mit Breitenangabe
\newcolumntype{C}[1]{>{\Centering\arraybackslash}p{#1}} % zentriert mit Breitenangabe
\usepackage{enumitem}
\usepackage{float}
\usepackage{pdflscape}
\usepackage{longtable}

\usepackage[all]{nowidow}

\AtBeginDocument{%
  \providecommand\BibTeX{{%
    \normalfont B\kern-0.5em{\scshape i\kern-0.25em b}\kern-0.8em\TeX}}}

\setcopyright{acmcopyright}
\copyrightyear{2020}
\acmYear{2020}
\acmDOI{10.1145/1122445.1122456}

\begin{document}

%%
%% The "title" command has an optional parameter,
%% allowing the author to define a "short title" to be used in page headers.
\title{A Survey on Human Machine Interaction in Industry 4.0}

%%
%% The "author" command and its associated commands are used to define
%% the authors and their affiliations.
%% Of note is the shared affiliation of the first two authors, and the
%% "authornote" and "authornotemark" commands
%% used to denote shared contribution to the research.
\author{Christian Krupitzer}
\email{christian.krupitzer@uni-wuerzburg.de}
\orcid{0000-0002-7275-0738}
\author{Veronika Lesch}
\email{veronika.lesch@uni-wuerzburg.de}
\orcid{0000-0001-7481-4099}
\author{Marwin Z\"ufle}
\email{marwin.zuefle@uni-wuerzburg.de}
\orcid{0000-0002-6620-9152}
\author{Samuel Kounev}
\email{samuel.kounev@uni-wuerzburg.de}
\orcid{0000-0001-9742-2063}
\affiliation{%
  \institution{University of W\"urzburg}
  \streetaddress{Am Hubland}
  \city{W\"urzburg}
  \country{Germany}
  \postcode{97074}
}
\author{Sebastian M\"uller}
\email{mueller.sebastian12@gmx.de}
\orcid{}
\author{Janick Edinger}
\email{janick.edinger@uni-mannheim.de}
\orcid{}
\author{Christian Becker}
\email{chrstian.becker@uni-wuerzburg.de}
\orcid{}
\affiliation{%
  \institution{University of Mannheim}
  \streetaddress{Schloss}
  \city{Mannheim}
  \country{Germany}
  \postcode{68161}
}
\author{Alexander Lemken}
\email{lemken@ioxp.de}
\orcid{}
\affiliation{%
  \institution{ioxp GmbH}
  \streetaddress{Julius-Hatry-Stra{\ss}e 1}
  \city{Mannheim}
  \country{Germany}
  \postcode{68163}
}
\author{Dominik Sch\"afer}
\email{Dominik.Schaefer@syntax.com}
\orcid{}
\affiliation{%
  \institution{Syntax Systems GmbH}
  \streetaddress{H\"ohnerweg 2-4}
  \city{Weinheim}
  \country{Germany}
  \postcode{69469}
}

%\author{Charles Palmer}
%\affiliation{%
%  \institution{Palmer Research Laboratories}
%  \streetaddress{8600 Datapoint Drive}
%  \city{San Antonio}
%  \state{Texas}
%  \postcode{78229}}
%\email{cpalmer@prl.com}

%%
%% By default, the full list of authors will be used in the page
%% headers. Often, this list is too long, and will overlap
%% other information printed in the page headers. This command allows
%% the author to define a more concise list
%% of authors' names for this purpose.
\renewcommand{\shortauthors}{Krupitzer et al.}

%%
%% The abstract is a short summary of the work to be presented in the
%% article.
\begin{abstract}
Industry 4.0 or Industrial IoT both describe new paradigms for seamless interaction between human and machines.
Both concepts rely on intelligent, inter-connected cyber-physical production systems that are able to control the process flow of the industrial production.
As those machines take many decisions autonomously and further interact with production and manufacturing planning systems, the integration of human users require new paradigms. 
In this paper, we provide an analysis of the current state-of-the-art in human machine interaction in the Industry 4.0 domain.
We focus on new paradigms that integrate the application of augmented and virtual reality technology.
Based on our analysis, we further provide a discussion of research challenges.
\end{abstract}

%%
%% The code below is generated by the tool at http://dl.acm.org/ccs.cfm.
%% Please copy and paste the code instead of the example below.
%%
\begin{CCSXML}
<ccs2012>
   <concept>
       <concept_id>10003120.10003121</concept_id>
       <concept_desc>Human-centered computing~Human computer interaction (HCI)</concept_desc>
       <concept_significance>500</concept_significance>
       </concept>
   <concept>
       <concept_id>10010405.10010481.10010482</concept_id>
       <concept_desc>Applied computing~Industry and manufacturing</concept_desc>
       <concept_significance>300</concept_significance>
       </concept>
    <concept>
       <concept_id>10011007.10011074.10011075</concept_id>
       <concept_desc>Software and its engineering~Designing software</concept_desc>
       <concept_significance>300</concept_significance>
       </concept>
 </ccs2012>
\end{CCSXML}

\ccsdesc[500]{Human-centered computing~Human computer interaction (HCI)}
\ccsdesc[300]{Applied computing~Industry and manufacturing}
\ccsdesc[300]{Software and its engineering~Designing software}

%%
%% Keywords. The author(s) should pick words that accurately describe
%% the work being presented. Separate the keywords with commas.
\keywords{human machine interaction, augmented / virtual reality, Industry 4.0, Industrial IoT}

%%
%% This command processes the author and affiliation and title
%% information and builds the first part of the formatted document.
\maketitle

\section{Introduction}
\label{chap:introduction}

The world has witnessed three industrial revolutions since the late eighteenth century which all brought major leaps forward in the efficiency and productivity of industrial activities and are largely acknowledged as historical facts, at least in Europe.

While the first and second industrial revolution were characterized by mechanization based on the invention of the steam engine and electrification of production processes, respectively, the third industrial revolution was induced by the introduction of computerized, thus automated, processes into manufacturing \cite{Kagermann2015}. 

What had been less agreed upon in the scientific community and among practitioners until recently is the question whether manufacturing is currently subject to another disruptive revolution, for the first time not only identified ex post. A distinctive feature of this alleged fourth industrial revolution, commonly referred to as \ac{I4.0}, is the integration of two formerly opposing paradigms of industrial activities, the so-called economies of scale and economies of scope, by means of mass production of individually customized products \cite{Wang2017,Shrouf2014}.

Opponents of the notion of another actual revolution advocate the view that \ac{I4.0} can be seen as a step-wise evolution based upon characteristic technologies of the third industrial revolution's computer-integrated manufacturing rather than representing a stand-alone revolution following a specific scientific break-through \cite{Ruttimann2016}. Besides, the true magnitude and implications of \ac{I4.0}, accused of being subject to predictive and exaggerating marketing efforts, will only be verifiable in retrospect \cite{Drath2014}. 

However, despite existing skepticism towards the revolutionary character and degree of maturity of \ac{I4.0} activities and enabling technologies, in the most recent years \ac{I4.0} has largely been established and accepted as a synonym for a fourth industrial revolution, either ongoing or on the verge of inception. This observation bases on a substantial amount of publications from information and communications technology and manufacturing science. The main reason for acknowledging \ac{I4.0} as an actual revolution in the industrial sector is the magnitude of economic effects and process-related implications it is expected to exert on the manufacturing industry, comparable to those of the preceding industrial revolutions \cite{Kagermann2015,Pereira2017}.

\subsection{Problem Statement and Motivation}
\label{sec:problem_motivation}

\textbf{Problem Statement.} Accepting the predominant classification of \ac{I4.0} as a revolutionizing development for the industrial sector, its relevance and significance for advanced economies is obvious. This holds true for Germany in particular, considering the fact that the German industrial sector is still comparably large and a crucial pillar of German economic power, having contributed 26.1\% of the country's overall gross value added in 2017 \cite[p.61]{StatistischesBundesamtDestatis2018}. 

At the same time, in light of modern globalized business and value chains, any corporate entity or even industry branch is at risk of losing market share and ultimately its raison d'\^{e}tre if it fails to successfully manage its cost structure and stay competitive. While the German economy, supported by its broad network of small and medium-sized enterprises, certainly represents a particular case regarding the significance of the industrial sector for the overall economy, advanced economies with a strong production industry, in general, face this challenge. Taking as given their disadvantage as opposed to competing emerging economies with respect to the general level of labor cost, competitiveness based on industrial activities of similar structure and nature seems out of reach. 

Instead, a promising resort might be found in the implementation of advanced automation solutions for manufacturing. The rationale behind said strategy lies in the hope for increased productivity induced by the application of superior technology, compensating for the disadvantage regarding the costs per piece. On top of that, the disadvantage in the cost structure might be reduced by the opportunity to implement less human labor-intensive operations provided by the increased degree of process automation. 

Despite the apparent economic appeal of such an approach, there is another crucial sphere to be considered in an evaluation of the overall problem and potential methods of resolution. In modern informed societies, both from a political and economic or managerial point of view, it is imperative to secure social sustainability when promoting such a far-reaching agenda.

Returning to the particularities of the German economic structure, a phenomenon closely linked to the size of its industrial sector is the large share of the active labor force employed in this sector. According to the German Federal Statistical Office (2018), the industrial sector employed 19.9\% of all employees in Germany in 2017 \cite[p.70]{StatistischesBundesamtDestatis2018}. Obviously, a large cut in jobs in the manufacturing industry due to increasing automation of formerly human tasks would have wide-ranging implications for the German economy and particularly the labor market and the affected individuals. 

On the other hand, securing competitiveness of domestic industrial activities, as indicated above, seems indispensable as well if a substantial share of jobs in the sector is supposed to be preserved. Hence, at first sight, decision-makers in businesses and government authorities seem to be running into the problem and dilemma of equipping manufacturers with appropriate tools, in particular technologies, for fierce competition while, at the same time, determining a tolerable amount and degree of process automation in order to preserve jobs in one of the most significant employment sectors.

\textbf{Motivation.} Fortunately, the manufacturing industry is currently evolving away from the notion that productivity enhancements through technological progress and sustainable human employment are mutually exclusive, instead moving towards the integration of both under the fourth industrial revolution. 

In particular, a central characteristic of \ac{I4.0} activities is supposed to be the integration of human operators into modern and advanced production processes, captured under the paradigm of human-centered automation in manufacturing \cite{Nelles2016}. This means that future competitiveness is not supposed to be secured only via superiority in automation-based productivity but rather through offering enhanced value to customers by providing individual product customization. For that purpose, meaningful integration of the strengths of both human and machine entity is supposed to enable an increase in manufacturing flexibility \cite{Brettel2014}. 

Thus, successful collaboration and interaction of humans with diverse and potentially innovative technological hardware and software components, i.e. machines, with the aim of achieving human-machine symbiosis \cite{Romero2016a} will gain enormously in importance in various facets of industrial production. Nonetheless, so far, and to the best of the author's knowledge, no comprehensive survey on \ac{HMI} in \ac{I4.0} and an attempt at a structural description and capture of the topic exists in research. The lack of said type of study, in combination with the topic's relevance for successful implementation of \ac{I4.0} activities and, in turn, for ensuring future competitiveness and employment in the industrial sector, provides the motivation for this paper. 

\subsection{Research Questions}
\label{sec:research_questions}

In order to fill the mentioned gap in research regarding a comprehensive study of the state of the art in \ac{HMI} related to \ac{I4.0} operations, this paper attempts to provide a wide-ranging overview and, in particular, a structural ordering of the different facets of \ac{HMI} in \ac{I4.0}. This kind of structured description of the overall topic, taking into account different perspectives and sub-components, can help distinguishing and integrating diverse research streams as well as foci and opinions of scientists. Such integration provides a sound basis for a comprehensive understanding of the topic's different facets and, consequently, potential orientation concerning open research issues to be addressed in the future. 

Therefore, this paper aims to analyze and find answers to three consecutive high-level research questions guiding the procedure and remaining structure of the paper. While the first research question, \ac{RQ1}, features a more qualitative character, the second and third research question, \ac{RQ2} and \ac{RQ3}, aim to motivate a more quantitative analysis based upon potential insights gained from answering \ac{RQ1}. 

In particular, \ac{RQ1} raises the question what the current state of the art in \ac{HMI} research in the area of \ac{I4.0} is. In the course of answering \ac{RQ1}, a comprehensive taxonomy of the topic is supposed to be developed, capturing and systematically structuring the most important facets of the topic. 

Relating to different elements of this taxonomy, \ac{RQ2} and \ac{RQ3} ask what the main foci and identifiable patterns in research related to \ac{HMI} in the \ac{I4.0} area are, respectively. On a more detailed level, \ac{RQ2} and \ac{RQ3} are split into several sub-questions specifying the respective focus in the identification of potential focal points and patterns in related research. The three sub-questions of \ac{RQ2} ask what the focal points in research related to the human, machine, or interaction aspect in \ac{I4.0}-related \ac{HMI} are, respectively. The sub-questions of \ac{RQ3}, on the other hand, focus on the identification of underlying patterns in research related to the topic and ask what kinds of associations can be identified in the data on related research and how related research articles can be classified with regard to individual aspects of the topic, respectively. A detailed listing of the three main research questions and their sub-questions, if applicable, is provided in Table \ref{table_research_questions}. 

%\begin{table}[h]
%\begin{tabular}{p{0.5cm} p{5.8cm} p{0.6cm} p{5.9cm}}
%\multicolumn{2}{l}{{\small \textbf{Main Research Questions}}} & \multicolumn{2}{l}{{\small \textbf{Sub-Questions}}} \\ \hline 
%\rule{0pt}{13pt}{\scriptsize \ac{RQ1}} & {\scriptsize What is the current state of the art in \acl{HMI} research in the area of \acl{I4.0}?} & & \\ 
%{\scriptsize \ac{RQ2}} & {\scriptsize What are the main foci and patterns in research related to \acl{HMI} in the area of \acl{I4.0}?} & {\scriptsize RQ2.1} & {\scriptsize What are the focal points in research related to the human aspect in \ac{I4.0}-related \ac{HMI}?} \\
% & & {\scriptsize RQ2.2} & {\scriptsize What are the focal points in research related to the machine aspect in \ac{I4.0}-related \ac{HMI}?} \\
% & & {\scriptsize RQ2.3} & {\scriptsize What are the focal points in research related to the interaction aspect in \ac{I4.0}-related \ac{HMI}?} \\
% & & {\scriptsize RQ2.4} & {\scriptsize What kinds of associations can be identified in the data on research related to \ac{HMI} in \ac{I4.0}?} \\
% & & {\scriptsize RQ2.5} & {\scriptsize How can research articles related to \ac{HMI} in \ac{I4.0} be classified with regard to individual aspects of the topic?} \\ 
%\end{tabular}
%\caption{Research Questions}
%\label{table_research_questions}
%\end{table}

\begin{table}[h]
	\begin{tabular}{p{1cm} p{12.3cm}}
		\multicolumn{2}{l}{{\textbf{Main Research Questions}}} \\ \hline 
		\rule{0pt}{15pt}{\small \ac{RQ1}} & {\small What is the current state of the art in \acl{HMI} research in the area of \acl{I4.0}?} \\ 
		\rule{0pt}{13pt}{\small \ac{RQ2}} & {\small What are the main foci in research related to \acl{HMI} in the area of \acl{I4.0}?} \\ 
		\rule{0pt}{13pt}{\small \ac{RQ3}} & {\small What are the patterns which can be identified in research related to \acl{HMI} in the area of \acl{I4.0}?} \\
		& \\
		\multicolumn{2}{l}{{\textbf{Sub-Questions}}} \\ \hline
		\rule{0pt}{15pt}{\small RQ2.1} & {\small What are the focal points in research related to the human aspect in \acl{I4.0}-related \acl{HMI}?} \\
		\rule{0pt}{13pt}{\small RQ2.2} & {\small What are the focal points in research related to the machine aspect in \acl{I4.0}-related \acl{HMI}?} \\
		\rule{0pt}{13pt}{\small RQ2.3} & {\small What are the focal points in research related to the interaction aspect in \acl{I4.0}-related \acl{HMI}?} \\
		\rule{0pt}{21pt}{\small RQ3.1} & {\small What kinds of associations can be identified in the data on research related to \acl{HMI} in \acl{I4.0}?} \\
		\rule{0pt}{13pt}{\small RQ3.2} & {\small How can research articles related to \acl{HMI} in \acl{I4.0} be classified with regard to individual aspects of the topic?} \\ 
	\end{tabular}
	\caption{Research Questions}
	\label{table_research_questions}
\end{table}

\subsection{Structure}
\label{sec:structure_paper}

Having presented a short introduction to the topic of \ac{HMI} in \ac{I4.0} including an emphasis on the relevance of research and progress in both \ac{I4.0} overall and its sub-aspect of \ac{HMI}, the paper continues in Section \ref{chap:foundations} with a definition and description of the relevant foundations for the overall paper. This includes a description of the fundamentals of \ac{HMI} and of two fundamental technologies for \ac{I4.0} activities (\acp{CPS} and the \ac{IoT}), followed by a detailed clarification of the meaning and scope of \ac{I4.0} as well as a description of a data mining approach applied in the course of this paper. 

Before Section \ref{chap:methodology} covers and describes the methodological approach of a structured literature review implemented in this paper, Section \ref{chap:related_work} presents an overview over studies related to the type and topic of this paper. 

As the central parts and answers to the research questions of this paper, Section~\ref{chap:results_taxonomy} and Section~\ref{chap:results_tax_analysis} present the results of the applied methodology and implemented analyses, starting with a description of the developed taxonomy in Section \ref{chap:results_taxonomy} and followed by a quantitative analysis of the prevalence of different taxonomy elements in the literature sample and of patterns in the collected data in Section~\ref{chap:results_tax_analysis}. 

%In order to put those results into perspective, Section~\ref{chap:discussion} discusses their implications for various stakeholders as well as the remaining challenges in implementing \ac{I4.0}-related \ac{HMI} processes and clarifies the limitations of the paper and its results due to the study's scope and methodology. 

Finally, in Section~\ref{chap:conclusion}, a summary of the central propositions of the study, combined with concluding remarks, follows which leads to an outlook on opportunities for future research derived from the insights of this paper. 

\section{Background}
\label{chap:foundations}

Before presenting and describing the study's methodology and results in depth beginning in Section~\ref{chap:methodology}, the following section aims to lay the groundwork for a general and common understanding of \ac{HMI} and \ac{I4.0}, especially for readers not yet broadly familiar with the topic of this paper. 

After a brief introduction to the concept of \ac{HMI} and related technologies, the concepts of \acp{CPS} and the \ac{IoT} are addressed in order to lay the foundations for a following detailed description of the umbrella term \ac{I4.0}.

\subsection{Human-Machine Interaction and Related Technologies}
\label{sec:definition_hmi}

While early scientific efforts on \ac{HMI} or human-computer interaction had concentrated on fully controllable systems, research quickly turned the focus towards adaptive mechanisms due to the development of dynamic, highly complex human-machine systems \cite{Hoc2000}. Exceeding the role of mere information processing, automated systems have evolved into partly independent actors in dynamic situations and uncertain environments which has led to the requirement of introducing elements of human-to-human relationships into \ac{HMI} and human-machine collaboration \cite{Hoc2000}. This implies that, in complex human-machine systems, the human and machine agent can no longer be considered in isolation and should instead be regarded as a dynamic unit or team collaborating towards an overall task and aim including dynamic task allocation among participants \cite{Hoc2000}.

Enabling adaptability and dynamism of the human-machine system has required the abandonment of full system-side determinism in order to leave degrees of freedom for the automated system to react to situations which could not have been perfectly foreseen and modeled by system designers \cite{Hoc2000}. Thus, the human operator's responsibility evolves from full control to partial control and supervision in the interaction with machines \cite{Hoc2000}.

In order to secure that the human operator maintains an overview and understanding of the processes and dynamics in this complex human-machine system, the types and characteristics of implemented interfaces remain a central aspect in the examination of \ac{HMI} processes, especially regarding ergonomic properties like usability and transparency \cite{Hoc2000}. As modern and advanced instances of \acp{UI}, \ac{AR} and \ac{VR} will play an important role in the discussion of \acp{UI} for \ac{HMI} purposes in \ac{I4.0} in the course of this paper. Therefore, brief definitions of those technologies are provided.

According to Paelke, \ac{AR} applications are characterized by enhancing a real-world environment through its integration with computer-generated information \cite{Paelke2014}. Mainly, this implies adding digital objects and information to a human user's real-world view, enabling direct interaction with information which has a direct spatial relation to the real environment and is more readily interpretable in its environmental context \cite{Paelke2014}.

In contrast, users of \ac{VR} applications do not necessarily maintain a view of a real-world scenery but are instead immersed in an entirely computer-generated and digital world offering interaction possibilities with components of the artificial environment \cite{Michalos2016}. Thus, both \ac{AR} and \ac{VR} enable the mediation of information in a spatial relation to the environmental context \cite{Paelke2014}, but \ac{AR} relies on a basis of data from a real-world scenery to be enhanced digitally while \ac{VR} applications are entirely based on computer-generated content \cite{Ehmann2018}.

\subsection{Cyber-Physical Systems and the Internet of Things}
\label{sec:cps_iot}

Representing main enablers leading to the development of \ac{I4.0} \cite{Wittenberg2016}, the concepts of \acp{CPS} in manufacturing and the \ac{IoT} are briefly introduced in this section. 

The concept of a \ac{CPS} has been known for more than a decade now, characterized primarily by its feature of integrating computational processes with the corresponding physical processes, i.e. synchronizing physical and cyber world. This means that physical processes are constantly monitored and controlled in the cyber space so that computation has a direct manipulating impact on real-world operations while, at the same time, feedback from physical processes affects computational procedures \cite{Lee2008}. 

While conceivable application scenarios for \acp{CPS} are numerous and diverse, the technology has received enormous attention in manufacturing research recently, especially regarding \ac{I4.0} applications. Therefore, \acp{CPS} deployed in a manufacturing context, mainly machines and production modules, are often referred to as \acp{CPPS}. \acp{CPPS} are characterized by their capabilities to collect and exchange data and information autonomously, to activate actual physical processes, and to perform independent mutual control via their embedded local intelligence \cite{Weyer2015,Assunta2017}.

Obviously, self-organization and mutual control among various instances of \acp{CPPS} requires context-awareness and communication channels connecting all involved entities which is provided by the ability to sense the environment in real time and the installation in a networked infrastructure \cite{Zhou2015,Weyer2015}.

Said interconnection among \acp{CPPS} represents an important instance of another fundamental pillar of \ac{I4.0} activities \cite{Wittenberg2016,Kagermann2015}, namely the \ac{IoT}. The \ac{IoT} can be considered as representing an extension to the well-known Internet by adding a huge amount of ending nodes to the network. What is special about those nodes added under the \ac{IoT} is that they are regularly small-scale computers regarding both the functionality and computing power they provide as well as the amount of energy they consume \cite{Fleisch2010}. This coincides with the above description of \acp{CPPS} added to the \ac{IoT} network which feature small-scale embedded sensors and processing units capable of serving a specific purpose, e.g. capturing and communicating operational and environmental data \cite{Hofmann2017}. 

Consequently, the \ac{IoT} represents a network infrastructure where any conceivable physical item, i.e. \texttt{thing}, can be embedded with computational intelligence featuring a connection to the Internet \cite{Fleisch2010}. With respect to \ac{HMI} processes in industrial operations, this implies that machinery and products of a \ac{CPS}-type offer a new quality of information transparency and availability by being able to provide data to an authorized entity at any time via the Internet.

\subsection{Industry 4.0}
\label{sec:industry_40}

Having introduced important technological foundations for a successful implementation of \ac{I4.0} activities, the question arises what the term \texttt{Industry 4.0} actually comprises. Therefore, the following subsections define the meaning and scope of \ac{I4.0} by first explaining the origin of the term and its evolution towards a synonym for a fourth industrial revolution and then introducing paradigms used in literature to illustrate the scope and concept of \ac{I4.0} operations.

\subsubsection{Origin and Evolution of the Term}
\label{subsec:i40_origin}

Interestingly, \texttt{\acl{I4.0}}, or rather the German equivalent \texttt{Industrie 4.0}, is not a term established by scholars in scientific publications. Instead, towards the end of 2011, the German federal government adopted the Industrie 4.0 project initiated a short while before by the German Industry-Science Research Alliance and made it an integral part of their high-tech strategy 2020 \cite{Kagermann2013}. In 2011, the term \texttt{Industrie 4.0} was also first introduced to the general public and the scientific community in the course of the Hannover Fair \cite{Vogel-Heuser2016a,Sanders2016}. 

Soon after, the concept and ideas of Industrie 4.0 found their way into scientific research and a growing number of studies were published dedicated towards different aspects of the topic. Indeed, some of the early contributions towards \ac{I4.0} actually even used the German instead of the English version of the term (cf.~\cite{Schuh2014a,Posada2015,Shafiq2015,Wan2015,Wang2016b}). As indicated in the Introduction, only over time, \ac{I4.0} evolved into a stand-alone term representing a synonym for future smart manufacturing and a fourth industrial revolution.

Nonetheless, the label \ac{I4.0} for innovative concepts of \ac{CPS}- and \ac{IoT}-integrated smart manufacturing is not unique in research activities from different areas. Particularly in the U.S., scientific publications frequently address either the so-called Industrial Internet, Industrial Internet of Things \cite{Wang2016c}, or simply smart manufacturing \cite{Thoben2017}. Besides, there are numerous national and regional initiatives similar to the German strategic initiative of Industrie 4.0, e.g. in the U.S. \cite{Posada2015,Jeschke2017}, China \cite{Lu2017}, Japan \cite{Wang2015}, South Korea \cite{Sung2018}, France, and the Basque Country in Spain \cite{Posada2015}.

\subsubsection{Paradigms}
\label{subsec:i40_paradigms}

After the clarification of the origin and evolution of the term itself, the following subsections are supposed to define and illustrate the fundamental pillars and constituents of the \ac{I4.0} concept. For that purpose, they present central paradigms of \ac{I4.0}, as identified and elaborated on in related scientific literature, in order to give the reader an overview over main features of \ac{I4.0} operations.

\paragraph{Actors in Future Manufacturing}
\label{subsubsec:i40_par_actors}

As indicated in Section \ref{sec:problem_motivation}, a central characteristic of manufacturing activities adhering to \ac{I4.0} is the integration of human factors into operations despite an advanced degree of process automation. This kind of human-centered perspective on smart manufacturing activities is reflected in the work of Weyer et al. who define three different types of actors in manufacturing as a central paradigm of future \ac{I4.0} activities. These actors are represented, on the one hand, by products and machinery turning into smart products and smart machines reflecting increased autonomy and automation under \ac{I4.0}, as well as, on the other hand, by augmented operators highlighting the significance of human involvement to ensure smooth running of operations \cite{Weyer2015}.

\textbf{Smart Products.} According to the authors of \cite{Weyer2015}, products under \ac{I4.0} evolve from passive objects, upon which actions are performed, into active subjects crucial to the implementation of decentralized process control. In that sense, every product is embedded with individual memory early on in the production process enabling products to self-orchestrate their assembly process. This means that unfinished products carry information about the remaining required process steps so that they can, e.g., order the specific service or part needed for their next step of assembly from production resources with open capacity \cite{Weyer2015}.

\textbf{Smart Machines.} The second type of central agents in \ac{I4.0} is represented by smart machines implying that modern machinery turns into \acp{CPPS}, thus featuring characteristics described in Section \ref{sec:cps_iot}, like context-awareness, self-organization, as well as mutual control and communication. Furthermore, in order to increase flexibility and adaptability of the overall production system, smart machines allow for a fast and convenient integration of a new \ac{CPPS} into an existing production cell, enabling a plug-and-play solution instead of a lengthy ramp-up phase for newly integrated manufacturing equipment \cite{Weyer2015}.

Despite those advanced capabilities and opportunities offered by smart products and especially machines, the complexity of modern, highly automated production processes implies that the human factor in production cannot be eliminated without restraining the overall system's operability and functionality \cite{Zamfirescu2014}. 

\textbf{Augmented Operators.} This is why Weyer et al. define a third type of central actor in \ac{I4.0} activities which is the  augmented operator. The distinct feature of this vivid human agent compared to its counterparts is the inherent adaptability of the human entity, making it the most flexible part of the entire operations. Assuming this role, the human is supposed to take responsibility of various crucial tasks to complement the overall, still highly automated system \cite{Weyer2015}. 

Yet, human operators in \ac{I4.0} are not supposed to rely solely on their inherent capabilities. Instead, in performing their tasks, they receive support by appropriate technologies in order to exploit their own full potential \cite{Weyer2015}. The coherent consequence of this interplay of human and machine actors in the course of manufacturing activities is the necessity for purposeful and successful \ac{HMI} in order to guarantee corporate success in \ac{I4.0}, leading to the topic of this paper.

\paragraph{Design Principles for Industry 4.0 Activities}
\label{subsubsec:i40_par_design}

Representing a suitable conclusion and integration of the technological and process-related characteristics of \ac{I4.0} introduced so far, this subsection provides an overview over design principles for \ac{I4.0} operations, as identified and defined in scientific literature on the anticipated fourth industrial revolution.

Based on a survey on the German \texttt{Industrie 4.0} combined with a nominal group workshop with corporate experts, Hermann et al. develop a framework of four major design principles for \ac{I4.0} activities. These are technical assistance, interconnection, information transparency, and decentralized decisions \cite{Hermann2016}. 

Practically, the first one of these design principles directly refers to the human role of an augmented operator in \ac{I4.0} activities, as introduced in the preceding section. To be more precise, technical assistance as a guideline for designing \ac{I4.0} operations suggests to provide both virtual and physical support for human workers in smart manufacturing environments. Virtual assistance implies support for the human operator in maintaining an overview and understanding of ongoing activities and processes via appropriate data and information visualization. Physical assistance, on the other hand, can be provided, for instance, by co-working robots performing physically demanding tasks \cite{Hermann2016}.

Interconnection hints towards the necessity to comprehensively connect production factors and stakeholders like manufacturing equipment and people via the Internet in order to enable seamless collaboration and information sharing. This, in turn, requires the establishment of common standards for communication and an appropriate level of cyber security when designing such a system \cite{Hermann2016}. 

This type of interconnected design of \ac{I4.0} activities directly leads to the remaining two design principles of information transparency and decentralized decisions. The ubiquitous connectivity of manufacturing equipment and employees enables the aggregation of raw data from the physical world to higher-value context information in the cyber space where it is made available transparently and comprehensibly to the appropriate and authorized entities. Finally, providing \acp{CPPS} and employees with such aggregated and context-related information tailored to their specific situation enables decentralized autonomy for those actors in decision-making and process control \cite{Hermann2016}.

\section{Related Work}
\label{chap:related_work}

After laying the foundations for the description of the actual in-depth analysis in the course of this paper, which will follow in the upcoming sections, an overview over related work and a differentiation of related studies from this paper are provided in order to justify the relevance of this study and to give the reader an idea of ongoing research activities in the field of \ac{I4.0}. Taking into account the type of study this paper represents, surveys on \ac{I4.0}-related topics are considered as relevant related scientific work. 

On the most general level, existing published surveys on \ac{I4.0} can be divided into two main clusters. The first group of papers includes surveys on aspects and subtopics of \ac{I4.0} other than \ac{HMI} while the second cluster comprises studies which do have a focus on \ac{HMI} in an \ac{I4.0} environment. However, with regard to a differentiation from this paper, those surveys investigate some specific partial aspect instead of aiming for a comprehensive description and classification of \ac{HMI} in \ac{I4.0}. For a more detailed description of the two clusters of related work, the following sections provide an overview over \ac{I4.0}-related surveys with and without a focus on \ac{HMI} aspects.

\subsection{Surveys on Industry 4.0 without a Focus on Human-Machine Interaction}
\label{sec:surveys_no_hmi}

Considering the fact that \ac{I4.0} has been an extensively covered topic in industrial research for the last couple of years, it is of little surprise that numerous surveys on some of its elements can be found. Among the related surveys identified, those examining different aspects compared to this paper often have a fairly technological focus. 

In their general study on \ac{I4.0} research, Saucedo-Mart{\'{i}}nez et al. identify nine central baseline technologies for \ac{I4.0} activities and cluster scientific publications included in their sample according to those technological categories \cite{Saucedo-Martinez2018}. Those baseline technologies are Big Data and analytics, autonomous robots, simulation, horizontal and vertical system integration, the industrial \ac{IoT}, cyber security, additive manufacturing, \ac{AR}, and the cloud. In order to determine more detailed insights on each of these technologies, the articles assigned to the respective category, representing research specialized on this type of technology, are analyzed in more depth (cf. \cite{Saucedo-Martinez2018}). 

In a similar approach, Kang et al. investigate the national initiatives for future manufacturing in Germany, the U.S., and South Korea with regard to the respective central technological components, cluster existing publications according to those technologies, and analyze the technologies based on the identified relevant literature \cite{Kang2016}. For Germany, those are \acp{CPS}, the \ac{IoT}, Big Data, cloud computing, and sensors, while for the U.S., smart energy and additive manufacturing, and for South Korea, smart energy, \ac{3D} printing, and holograms are added \cite{Kang2016}.

Brettel et al. conduct an extensive literature review based on an immense sample of 5911 articles from practice-oriented journals targeted at industrial executives, combined with expert interviews from the industrial and consulting sector \cite{Brettel2014}. In their structured literature review, they map research on \ac{I4.0} according to the topics of individualization of production, horizontal integration in collaborative networks, and end-to-end digital integration, as well as the respective sub-aspects. In order to evaluate the practical relevance of their findings, they conduct the mentioned structured interviews with experts from practice \cite{Brettel2014}.

Lu provides another general survey on \ac{I4.0}, clustering related literature according to the five categories of concept and perspectives of \ac{I4.0}, \acp{CPS}, key technologies, interoperability, and applications of \ac{I4.0} \cite{Lu2017}. In the analysis of each of the five categories, Lu applies a particular focus on interoperability aspects of \ac{I4.0}, developing a broad conceptual framework of the topic \cite{Lu2017}.

Preuveneers \& Ilie-Zudor, in turn, survey the general developments and trends in \ac{I4.0}, identifying remaining challenges to the implementation of such activities which offer opportunities for future research \cite{Preuveneers2017}. The challenges identified and to be further researched are ways of guaranteeing predictable system behavior, quality assurance for context-aware behavior, risks with shifting intelligence from operators to automated systems, and compliance regulations and legal implications.

Finally, Pfohl et al. analyze the determinants for the diffusion of technological innovations under an \ac{I4.0} scenario in their structured literature review \cite{Pfohl2017} while Hermann et al. survey the topic of design principles for \ac{I4.0} activities covered in corresponding research \cite{Hermann2016}\footnote{Recall Section \ref{subsubsec:i40_par_design} for more details on the insights on \ac{I4.0} design principles gained from the study of Hermann et al. \cite{Hermann2016}.}. The diffusion factors for technological innovations in \ac{I4.0} identified in the former study are the degree of maturity of the communication system, know-how of staff, industry-wide investment efficiency, and industry membership \cite{Pfohl2017}.

\subsection{Surveys on Industry 4.0 with a Focus on Human-Machine Interaction}
\label{sec:surveys_hmi}

A similar number of surveys covering specific facets of \ac{HMI} or human factors in \ac{I4.0} has been identified in the course of collecting relevant literature for this paper. The vast majority of these studies is specifically oriented towards either \ac{VR} or \ac{AR} applications, or both, within \ac{I4.0} operations and, thus, covers only a subset of this paper's scope.

B{\"{u}}ttner et al. conduct a survey on \ac{AR} and \ac{VR} applications in \ac{I4.0} manufacturing activities, more precisely on the available platform technologies and application areas, creating a small-scale design space for such Mixed Reality applications in manufacturing \cite{Buttner2017}. This design space differentiates among four general application scenarios and four types of Mixed Reality technology platforms available for application, respectively. The application scenarios are manufacturing, logistics, maintenance, or training while the available platforms comprise mobile devices (\ac{AR}), projection (\ac{AR}), and \acp{HMD} (\ac{AR} or \ac{VR}).

Dini \& Dalle Mura and Wang et al. both present surveys on \ac{AR} applications, however not restricted to \ac{I4.0}-related application scenarios \cite{Dini2015,Wang2016a}. Dini \& Dalle Mura investigate general commercial \ac{AR} applications including besides industrial scenarios also, among others, civil engineering. The specific \ac{AR} application scenarios which they examine based on related scientific literature are maintenance and repairing, inspection and diagnostics, training, safety, and machine setup \cite{Dini2015}. Wang et al., in turn, examine scientific research on \ac{AR} applications for assembly purposes from a time span of 26 years starting as early as 1990 and concentrating mainly on the period from 2005 until 2015. Thus, they extend the scope to many years before the advent of \ac{I4.0}-related initiatives and ideas. The major application purposes of \ac{AR} in assembly tasks which they investigate are assembly guidance, assembly design and planning, and assembly training \cite{Wang2016a}. 

Lukosch et al. provide a literature review on \ac{AR} applications with an even less specific focus on industrial deployment by examining the state of the art at the time in research on collaboration in \ac{AR} considering a wide range of possible application fields, the industrial sector being only one of those \cite{Lukosch2015}. As a result, they identify remaining research challenges relating to collaboration in \ac{AR} which are the identification of suitable application scenarios and interaction paradigms as well as an enhancement of the perceived presence and situational awareness of remote users \cite{Lukosch2015}.

Palmarini et al., in turn, conduct a structured literature review on different software development platforms and types of data visualization and hardware available for \ac{AR} applications in various maintenance scenarios \cite{Palmarini2017}. The aim and purpose of their study is to derive a generic guideline facilitating a firm's selection process of the appropriate type and design of \ac{AR} application, tailored to the specific type of maintenance activity at hand which the firm is planning to enhance utilizing \ac{AR} technology \cite{Palmarini2017}.

Lastly, Choi et al. and Turner et al. provide surveys on \ac{VR} technology in an industrial environment, the latter group of authors concentrating on a potential combination of \ac{VR} technology with discrete event simulation for scenario testing in \ac{I4.0} activities \cite{Turner2016,Choi2015}. Choi et al., on the other hand, present a survey on \ac{VR} applications in manufacturing, concentrating on potential contributions of \ac{VR} deployment in the development process for new products and deriving a mapping of different types of \ac{VR} technology towards the different steps of the product development process. Therefore, they consider applicability of various \ac{VR} technologies for the phases of concept development, system-level design, design of details, testing and refinement, and launch of production \cite{Choi2015}. 

Besides those surveys on \ac{AR} and \ac{VR} applications, a literature review by Hecklau et al. exists on the major challenges as well as skills and competencies needed for future employees under an \ac{I4.0} scenario \cite{Hecklau2016}. The authors utilize the insights from the literature analysis to structure the required skills according to different categories, based on which a competence model is created analyzing employees' levels of skills and competencies which will be particularly important in an \ac{I4.0} working environment. Those main categories for the competence model are technical, methodological, social, and personal competencies \cite{Hecklau2016}.

Finally, it is important to highlight that this outline of related work does not claim to be exhaustive. Instead, it is rather supposed to provide an overview and an idea of other existing scientific literature regarding the topic and thereby underline that, to the best of the author's knowledge, a study highly similar to this paper does not exist. This observation serves for justifying the conduct of this study and, at the same time, implicates the necessary importance and relevance of the paper.

\section{Methodological Approach for Defining and Analyzing a Taxonomy}
\label{chap:methodology}

Having laid the foundations for an in-depth analysis of \ac{HMI} in \ac{I4.0}, the following section serves the purpose of explaining the methodological approach taken in this paper in order to define and analyze a comprehensive taxonomy of the topic.

Considering the abundance of scientific contributions towards \ac{HMI} in \ac{I4.0} and the lack of a comprehensive study attempting to capture and structure the entirety of the topic's facets, the development of a taxonomy has been chosen as an appropriate tool to illustrate an integration of the topic's various sub-aspects.

\begin{figure}[hbt]
	\centering
	\includegraphics[width=\textwidth]{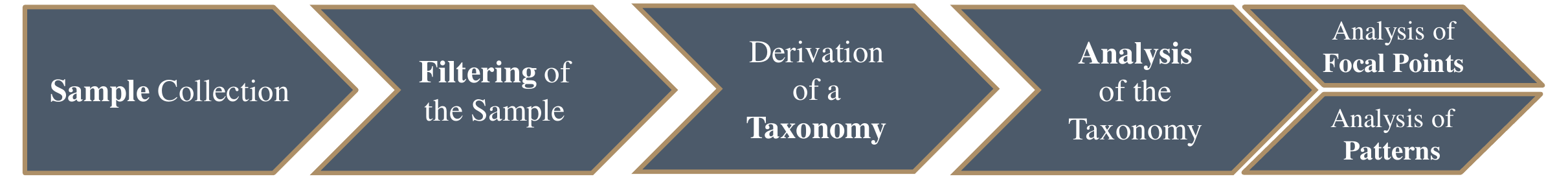}
	\caption{The Overall Approach Applied in the Course of this Survey}
	\label{figure:approach_paper}
\end{figure}

For this purpose, the methodological approach of a structured literature review in the form of a survey has been applied in order to identify all of the most relevant aspects of the topic and to ground the derivation of a taxonomy on a broad basis of profound scientific publications. The overall approach, illustrated in Figure \ref{figure:approach_paper}, comprises four major methodological steps which are the initial collection of a literature sample, the subsequent filtering of this collection, the derivation of a taxonomy based on the filtered final sample, and the analysis of this taxonomy. The latter step, in turn, comprises the analysis of focal points in research on the topic of this paper and of potential patterns in the corresponding literature.

The aim is to develop a taxonomy reflecting past and current scientific efforts in the field which can, consequently, serve as an orientation for related researchers concerning available contributions and opportunities for further research. In particular, the taxonomy aims to grant inspiration for novel perspectives on unexplored combinations and interactions of two or more different facets of \ac{HMI} in \ac{I4.0}. Therefore, based on a large-scale survey, the taxonomy is supposed to integrate various, possibly heterogeneous perspectives of different authors.

\subsection{Literature Search for Initial Sample Collection}
\label{sec:sample_collection}

In a first step towards developing a taxonomy, potential scientific literature representing candidates for the final sample of relevant publications had to be collected. 

For this purpose, an approach comparable to the search procedure in a survey by Lu \cite{Lu2017} was chosen which is based on Webster \& Watson \cite{Webster2002}. However, the steps of going forward and backward, i.e. of including both the articles citing the publications identified in the keyword search and the articles cited by the identified publications, have been omitted. 

Instead, the approach applied for sample collection includes the development of a rich set of keywords for literature search as well as the choice of keyword combinations to be used for search and of search platforms for implementation of the search. The subsequent steps comprise the actual conduct of several search rounds using different keyword combinations and search platforms and the corresponding decision which of the search results to include in the initial sample.

\begin{figure}[h]
	\centering
	\includegraphics[width=\textwidth]{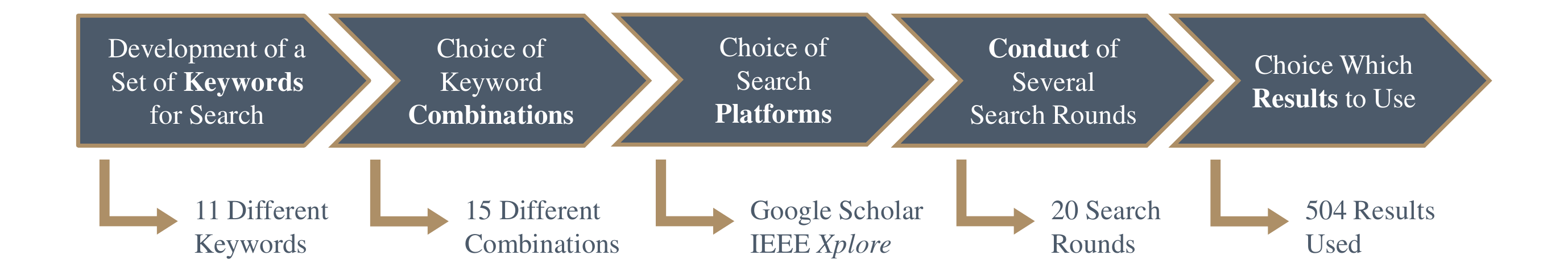}
	\caption{The Approach Applied for Collection of the Initial Sample}
	\label{figure:approach_sample_collection}
\end{figure}

The intention in relying on this approach, illustrated in Figure \ref{figure:approach_sample_collection}, instead of following Webster \& Watson \cite{Webster2002} in going backward and forward has been to minimize the risk of remaining within a specific scope of research determined by the direction of the starting articles, i.e. the starting points for going forward and backward. Instead, the aim has been to achieve a comprehensive sample of publications representing different research streams and scientific perspectives on the topic. Besides, as described further below, the applied approach has resulted in a final sample of substantial size, making a procedure of going backward and forward for every article in the sample impractical under the scope of this paper.

Regarding the identification of a set of keywords, considering the explicit focus of the paper on \ac{HMI} in an \ac{I4.0} environment, the rationale was to combine a diverse collection of keywords aiming at \ac{HMI} aspects with either \texttt{Industry 4.0} or the German \texttt{Industrie 4.0} occasionally used in scientific publications (see Section \ref{subsec:i40_origin}). This means that, for every search round, a specific keyword aiming at the \ac{HMI} aspect was combined with either \texttt{Industry 4.0} or \texttt{Industrie 4.0}. 

The overall set of keywords covering the \ac{HMI} aspect comprises \texttt{Human-Machine Interaction}, \texttt{HMI}, \texttt{Virtual Reality}, \texttt{VR}, \texttt{Augmented Reality}, \texttt{AR}, \texttt{Human Role}, \texttt{Human-Centered Design}, and \texttt{Human-Machine Collaboration} and is presented in Table \ref{table_keywords} together with the \ac{I4.0} keywords.

Regarding the choice of a platform or digital library for implementation of the search, Google Scholar was chosen as the initial search medium. The reason for deploying Scholar was to increase the chances of identifying literature from diverse research streams, publishing entities, and possibly even scientific disciplines for a sample covering multiple perspectives on \ac{HMI} in \ac{I4.0}.

\begin{table}[h]
	\begin{tabular}{p{5.3cm} p{2.2cm} p{5.8cm}}
		\multicolumn{2}{l}{\textbf{\small{Keywords Covering the \ac{HMI} Aspect}}} & \textbf{\small{Keywords for the \ac{I4.0} Aspect}} \\ \hline
		\small{Human-Machine Interaction} & \small{HMI} & \small{Industry 4.0} \\
		\small{Virtual Reality} & \small{VR} & \small{Industrie 4.0} \\
		\small{Augmented Reality} & \small{AR} & \\
		\small{Human-Centered Design} & \small{Human Role} & \\
		\small{Human-Machine Collaboration} & & \\
	\end{tabular}
	\caption{Keywords for Literature Search}
	\label{table_keywords}
\end{table}

However, the opening search round using the string \texttt{Human-Machine Interaction Industry 4.0} revealed the general observation that coverage in Google Scholar exceeds the aspired broadness of range by returning tremendous amounts of search results. Judging by titles, those results became ever less focused on topics related to \ac{HMI} in \ac{I4.0} the lower down the paper in the order of results returned. 

Besides, those results revealed that a significant amount of articles with more suitable titles stemmed from \ac{IEEE} sources which is why, for the fifteen subsequent search rounds, \ac{IEEE}'s digital library \ac{IEEE} \textit{Xplore} was used instead of Google Scholar. Thereby, the thematic focusing of search results was supposed to be enhanced.

As a matter of fact, search rounds on \ac{IEEE} \textit{Xplore} returned significantly less results than provided by Google Scholar. While the initial round on the latter platform had returned more than 9,000 results, the highest number of articles returned in a single round on \textit{Xplore} was 24. Therefore, all results returned by \textit{Xplore} have been included in the initial sample whereas Scholar results have only been taken into consideration up to a maximum of 200 results per search round.

Overall, a total of twenty search rounds has delivered an initial sample of 504 search results. As described above, the first round was conducted using Google Scholar while the following fifteen search rounds were implemented on \ac{IEEE} \textit{Xplore}. In order to further expand the sample of 323 search results after these first sixteen rounds and to avoid an overly narrow focus on \ac{IEEE} publications, the final four search rounds delivering the last 181 results for the initial sample were conducted on Google Scholar again.
%A comprehensive list of the twenty search rounds conducted including the respective number of results obtained for the initial sample can be found in Appendix \ref{chap:apx_sample_collection}.

\subsection{Filtering of the Sample}
\label{sec:sample_filtering}

\begin{figure}[h]
	\centering
	\includegraphics[width=9cm]{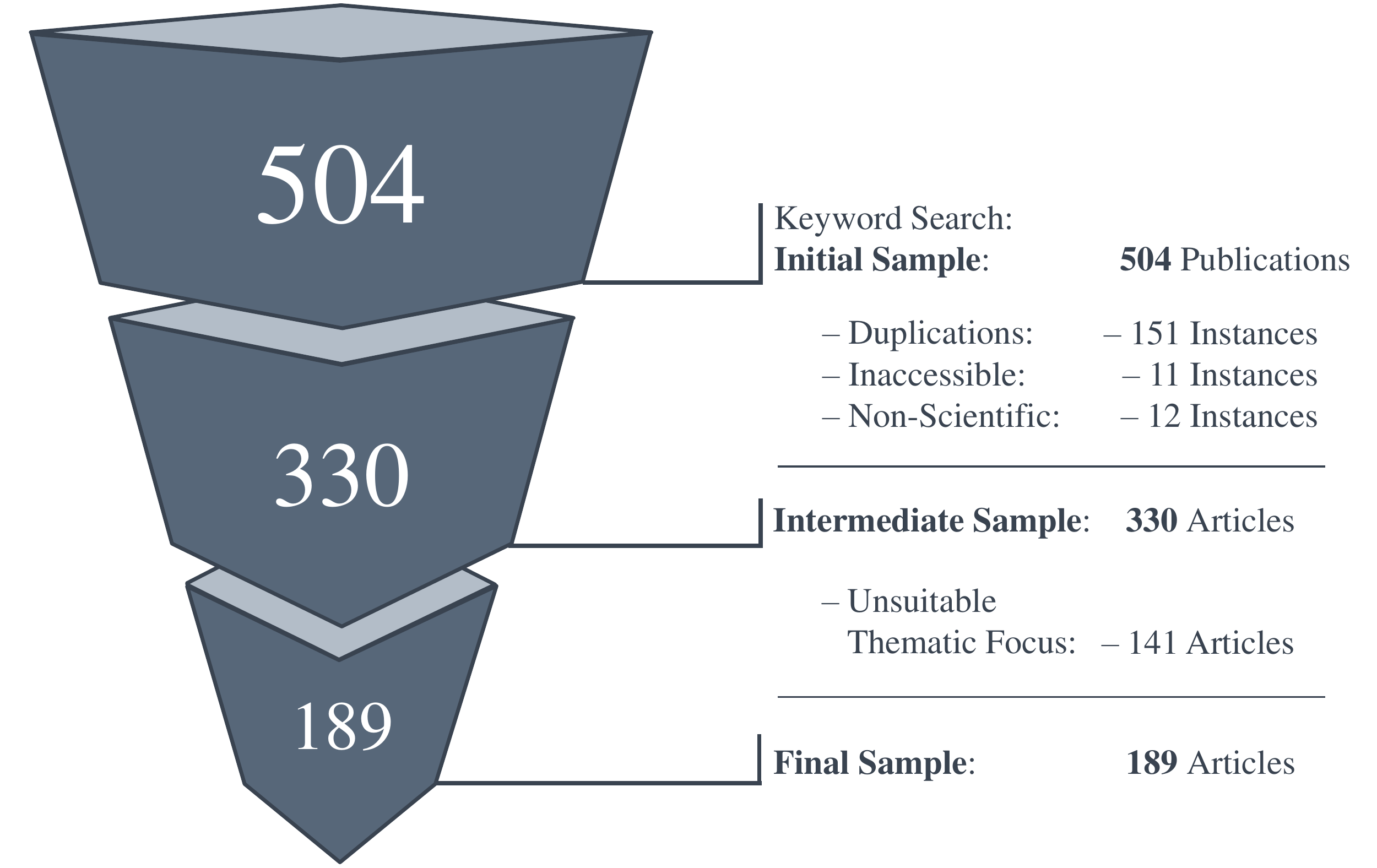}
	\caption{Process of Filtering the Initial Literature Sample}
	\label{figure:sample_filtering}
\end{figure}

Having collected a total of 504 search results as an initial sample for this survey, a crucial step in advance of developing a taxonomy of \ac{HMI} in \ac{I4.0} has been to sort out those instances of the sample which were either redundant, not available for access, represented non-scientific publications, or proved to be focused on topics irrelevant to the development of the taxonomy. By eliminating the first three types of results from the initial sample, an intermediate sample of 330 articles was generated before a detailed analysis and evaluation of content-related match of those articles with the research focus of this paper has led to an eventual elimination of another 141 articles, as illustrated in Figure \ref{figure:sample_filtering}. Thus, the final sample comprises 189 publications representing the basis for the development of a taxonomy of the topic.

In the following, this section describes the process of filtering the initial sample, as illustrated in total in Figure \ref{figure:sample_filtering}, stepwise and in more detail.

\subsubsection{Elimination of Redundant and Unsuitable Literature}
\label{subsec:sample_filtering_elimination}

A first step in filtering the initial collection of literature has been to identify all instances of the same publication occurring multiple times in this sample. In case of such duplications, only the article collected in the earliest respective search round has been retained for the intermediate sample. In rare cases, duplicated articles from later search rounds have been retained instead of an earlier collected instance if the publication type of the article collected later was of higher priority, in the order of journal article over conference proceeding over working paper.

Overall, among the 504 search results included in the initial sample, 353 distinct publications have been identified whereas each of the remaining and discarded 151 instances represented a duplication, either in the form of a direct copy, as the same work published via a different channel, or e.g. as the published German version of an English publication among the 353 distinct instances. 

In a next, straightforward step, the retained 353 publications have been filtered for those not available for a detailed analysis due to the publishing entity denying access to the full text of the article. In total, eleven publications and their full texts have been inaccessible and, thus, have been removed from the sample.

Finally, another twelve publications have been removed from the sample of distinct search results because they did not represent a type of article conforming to standards and requirements of scientific publications. Among those, five articles represented white papers published by corporate entities, mostly consulting firms, and another five were publicly available patents. Lastly, one search result in the form of a collection of presentation slides instead of an actual article and one published Master thesis were discarded. 

Altogether, removing 151 duplicated, eleven inaccessible, and twelve non-scientific publications from the initial sample of 504 search results led to an intermediate sample of 330 articles representing the basis for the in-depth analysis of literature contents towards the development of the taxonomy. 

In the course of reading and analyzing the contents of the articles included in the intermediate sample, 141 papers have been identified as focusing on topics irrelevant to the scope of this paper. Therefore, in a final step in filtering the sample of search results, they have been removed from the intermediate sample. 

In some cases, examining the title and abstract has sufficed to unequivocally determine that the respective article concentrated on topics irrelevant to this paper. In contrast, papers actually dedicated towards \ac{I4.0} have been read thoroughly before potentially deciding that they did not sufficiently cover aspects relevant to a taxonomy of \ac{HMI} in \ac{I4.0} in order to be included in the final sample. 

The following section describes the properties of the resulting final sample of 189 scientific publications which constitutes the basis for all of the following analyses conducted and presented in the course of this paper.

\subsubsection{Description of the Final Sample}
\label{subsec:sample_description_final}

The composition of the final sample can be analyzed with respect to different characteristics of the articles included. In particular, the final sample has been examined regarding the distribution of publication years and types as well as regions of origin among the total of 189 articles. In addition, the distribution of publication types has been scrutinized for conferences, (volumes of) specific journals, or books and book series more strongly represented within the sample. 

The first notable fact when examining the articles' publication years is that there are three instances from a period before the actual establishment of the term and the concept of \ac{I4.0}. The reason for including those articles into the final sample despite their age is the relevance and affiliation of their contents towards the focus of this paper, contributing insights concerning the cooperation of workers and robots in assembly lines in order to enhance flexibility \cite{Kruger2009}, a methodology for the analysis of quality in \ac{HMI} \cite{NolimoSolman2002}, and a very early and visionary concept of a human-machine interface implementing \ac{VR} aspects \cite{Shimoda1999}. 

Apart from this triple of articles, only scientific publications from after 2012 are included in the final sample, with the largest share of 72 articles (38.1\%) stemming from 2016. In general, the vast majority of publications stems from the period between 2015 and 2017, with another 22.2\% and 21.7\% of articles published in 2015 and 2017, respectively. 

\begin{figure}[h]
%	\begin{addmargin}{\dimexpr -\oddsidemargin-0.4in\relax}
		\begin{minipage}{0.49\textwidth}
			\includegraphics[width=\textwidth]{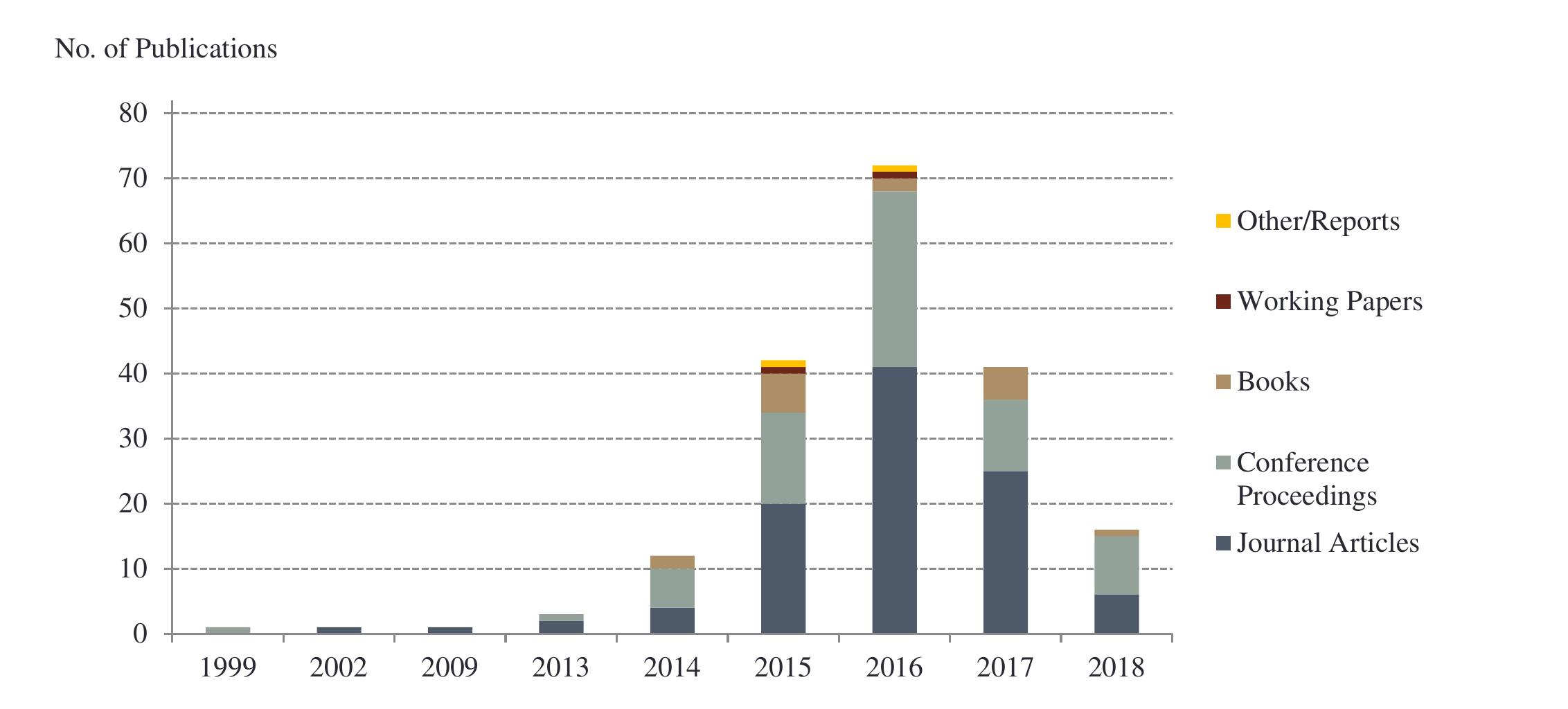}
		\end{minipage}
		\begin{minipage}{0.49\textwidth}
			\includegraphics[width=\textwidth]{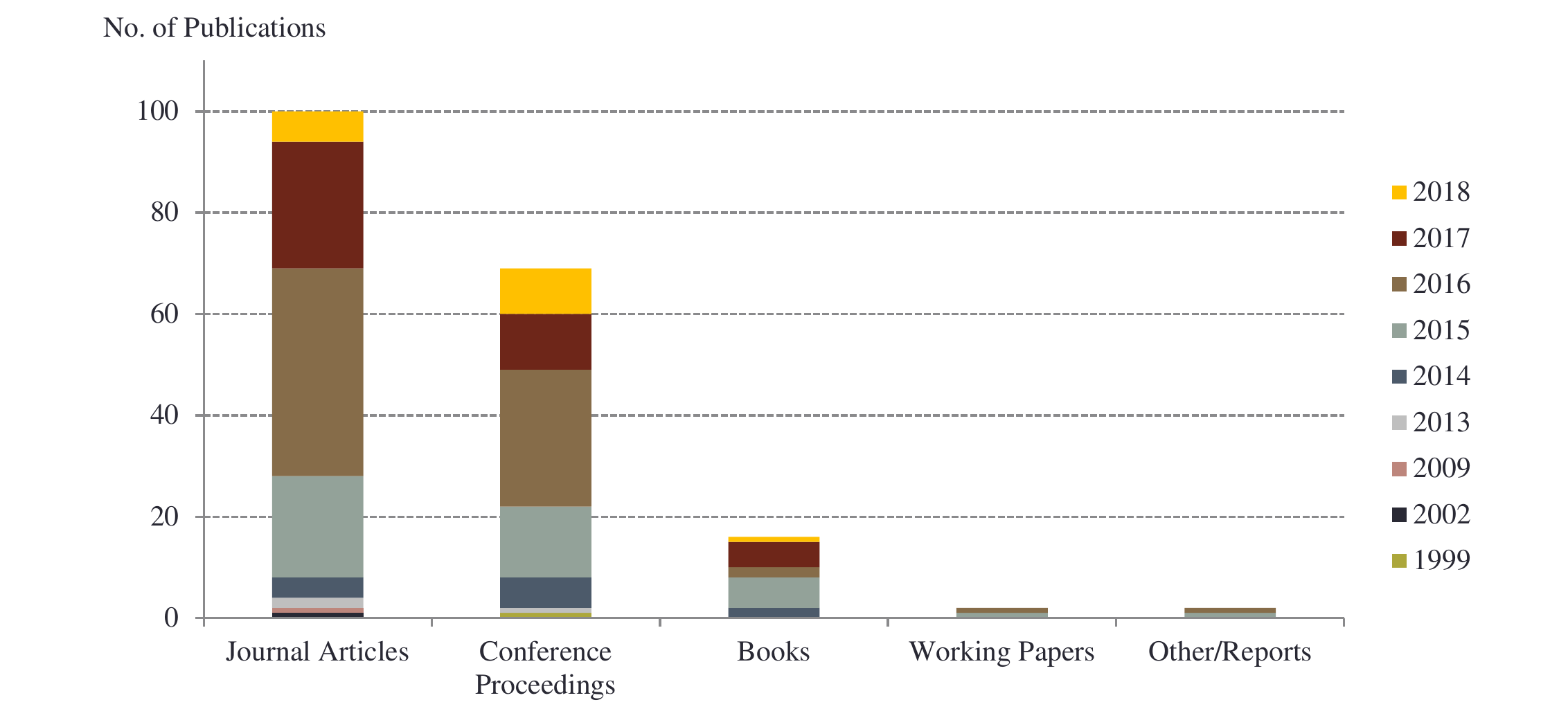}
		\end{minipage}
%	\end{addmargin}
	\caption{Distributions of Years and Types of Publication in the Final Sample}
	\label{figure:sample_distribution_years_types}
\end{figure}

Figure \ref{figure:sample_distribution_years_types} presents an overview of the overall distribution of years of publication in the final sample including information on the respective types of publication as well as the corresponding overview for the distribution of types of publication. Noticeable in this regard is the fact that, relating to the main period from 2015 to 2017, the proportion of journal articles among the sample instances is by far the highest while, for the year before and after that period, conference proceedings represent the highest share. 

%\begin{figure}[h]
%\includegraphics[width=13cm]{grafiken/distribution_publication_years.pdf}
%\caption{Distribution of Years of Publication in the Final Sample}
%\label{figure:sample_distribution_years}
%\end{figure}
%
%\begin{figure}[h]
%\includegraphics[width=13cm]{grafiken/distribution_publication_types.pdf}
%\caption{Distribution of Types of Publication in the Final Sample}
%\label{figure:sample_distribution_types}
%\end{figure}

The predominant role of journal articles also applies to the overall sample, as can be inferred from Figure \ref{figure:sample_distribution_years_types} as well. More than half of the sample papers were published in a scientific journal, the vast majority of which in the period from 2015 until 2017 and particularly in 2016, whereas conference proceedings make up a little more than a third of the sample and 8.5\% of the articles were published in the form or as part of a book. The years of publication concerning conference proceedings and books are a little more evenly distributed among the five-year period from 2014 to 2018 compared to journal articles. 

Besides, the final sample comprises two published working papers focusing on topics relevant to the scope of this paper and another couple of articles or reports published by the German political foundation Friedrich-Ebert-Stiftung also covering aspects relevant to this paper and authored by renowned German scientists.

\begin{table}[h]
%	\begin{addmargin}{\dimexpr -\oddsidemargin-0.3in\relax}
    \resizebox{\textwidth}{!}{
		\begin{tabular}{p{6.3cm} p{2.5cm} p{1.3cm} p{1.3cm} p{1.3cm} p{1.3cm}}
			\textbf{Scientific Journals} & & & & & \\ \hline
			{\small Journal} & {\scriptsize No. of Occurrences} & \multicolumn{4}{l}{{\scriptsize Vol.(No.) [No. of Occurrences]}} \\ \hline
			{\scriptsize Procedia CIRP} & \multicolumn{1}{c}{{\scriptsize 28x}} & \multicolumn{4}{l}{\scriptsize 17 [2x] \hspace*{0.07cm} 29 [1x] \hspace*{0.07cm} 30 [1x] \hspace*{0.07cm} 32 [1x] \hspace*{0.07cm} 40 [1x] \hspace*{0.07cm} 41 [4x]} \\
			& & \multicolumn{4}{l}{\scriptsize 44 [2x] \hspace*{0.07cm} 46 [1x] \hspace*{0.07cm} 53 [1x] \hspace*{0.07cm} 54 [5x] \hspace*{0.07cm} 55 [1x] \hspace*{0.07cm} 56 [2x]} \\
			& & \multicolumn{2}{l}{\scriptsize 57 [2x] \hspace*{0.07cm} 59 [1x] \hspace*{0.07cm} 63 [3x]} & \multicolumn{2}{l}{} \\ 
			{\scriptsize IEEE Access} & \multicolumn{1}{c}{{\scriptsize 6x}} & {\scriptsize 5 [1x]} & {\scriptsize 6 [5x]} & & \\  
			{\scriptsize Procedia Manufacturing} & \multicolumn{1}{c}{{\scriptsize 5x}} & {\scriptsize 1 [1x]} & {\scriptsize 9 [1x]} & {\scriptsize 11 [2x]} & {\scriptsize 13 [1x]} \\
			{\scriptsize CIRP Annals} & \multicolumn{1}{c}{{\scriptsize 3x}} & {\scriptsize 58(2) [1x]} & {\scriptsize 65(2) [2x]} & & \\
			{\scriptsize Computers in Industry} & \multicolumn{1}{c}{{\scriptsize 3x}} & {\scriptsize 65(2) [1x]} & {\scriptsize 83 [1x]} & {\scriptsize 89 [1x]} & \\
			{\scriptsize IEEE Computer Graphics and Appl.} & \multicolumn{1}{c}{{\scriptsize 3x}} & {\scriptsize 35(2) [3x]} & & & \\
			{\scriptsize IFAC-PapersOnLine} & \multicolumn{1}{c}{{\scriptsize 3x}} & {\scriptsize 48(3) [2x]} & \multicolumn{2}{l}{{\scriptsize 49(19) [1x]}} & \\
			{\scriptsize Procedia Engineering} & \multicolumn{1}{c}{{\scriptsize 3x}} & {\scriptsize 69 [1x]} & {\scriptsize 100 [1x]} & {\scriptsize 182 [1x]} & \\ \hline
			\hline
			\rule{0pt}{15pt}\textbf{Conference Proceedings} & & & & & \\ \hline
			{\small Conference} & {\scriptsize No. of Occurrences} & \multicolumn{4}{l}{{\scriptsize Conference Edition [No. of Occurrences]}} \\ \hline
			{\scriptsize IEEE Conference on Emerging Technologies \& Factory Automation} & \multicolumn{1}{c}{{\scriptsize 4x}} & \multicolumn{4}{l}{{\scriptsize 2014 19th [1x] \hspace*{0.07cm} 2015 20th [3x]}} \\
			{\scriptsize IEEE Int'l Conference on Industrial Informatics} & \multicolumn{1}{c}{{\scriptsize 4x}} & \multicolumn{4}{l}{{\scriptsize 2014 12th [1x] \hspace*{0.07cm} 2016 14th [2x] \hspace*{0.07cm} 2017 15th [1x]}} \\
			{\scriptsize Hawaii Int'l Conference on System Sciences} & \multicolumn{1}{c}{{\scriptsize 3x}} & \multicolumn{2}{l}{{\scriptsize 2016 49th [3x]}} & \multicolumn{2}{l}{} \\
			{\scriptsize IEEE Int'l Conference on Industrial Technology} & \multicolumn{1}{c}{{\scriptsize 3x}} & \multicolumn{4}{l}{{\scriptsize 2016 [2x] \hspace*{0.07cm} 2017 [1x]}} \\ \hline
			\hline
			\rule{0pt}{15pt}\textbf{Books} & & & & & \\ \hline
			{\small Book} & {\scriptsize No. of Occurrences} & & & & \\ \hline
			{\scriptsize Advances in Production Technology} & \multicolumn{1}{c}{{\scriptsize 3x}} & & & & \\
			{\scriptsize Service Orientation in Holonic and Multi-agent Manufacturing} & \multicolumn{1}{c}{{\scriptsize 2x}} & & & & \\ \hline
			\hline
			\rule{0pt}{15pt}\textbf{Book Series} & & & & & \\ \hline
			{\small Book Series} & {\scriptsize No. of Occurrences} & \multicolumn{4}{l}{{\scriptsize Volume [No. of Occurrences]}}\\ \hline
			{\scriptsize Lecture Notes in Production Engineering} & \multicolumn{1}{c}{{\scriptsize 3x}} & {\scriptsize n.a.} & & & \\
			{\scriptsize IFIP Advances in Inform. and Comm. Techn.} & \multicolumn{1}{c}{{\scriptsize 2x}} & {\scriptsize 488 [1x]} & {\scriptsize 513 [1x]} & & \\
			{\scriptsize Studies in Computational Intelligence} & \multicolumn{1}{c}{{\scriptsize 2x}} & {\scriptsize 594 [1x]} & {\scriptsize 640 [1x]} & & \\ \hline
		\end{tabular}}
	%\end{addmargin}
	\caption{Frequently Occurring Publication Mediums Among the Final Sample}
	\label{table_publication_mediums}
\end{table}

On a more detailed level, it is worth examining the list of actual journals, conference proceedings, and books included in the sample. In total, the final literature sample comprises articles from 44 different scientific journals, conference proceedings from 54 series of scientific conferences, and thirteen distinct published books, nine of which are instances of one of eight different book series. 

Table \ref{table_publication_mediums} presents an overview of all journals and conferences represented at least threefold and of books and book series represented at least twice within the final sample of this paper. This includes detailed information on the respective volume and number, if applicable, of the journal, edition of the conference, or volume of the book series as well as the exact number of distinct occurrences of these published instances within the final sample.

Finally, the countries of origin of the articles have been analyzed. More precisely, for every publication in the sample, the country where the affiliated organization of the author(s) is domiciled has been determined. For that purpose, in case of various authors with differing geographic provenience, the country of origin of the article has been decided based on majority votes. In case of a tie, the country of the corresponding author's affiliated organization has been chosen, whereas in case of a lack of information concerning the corresponding author, origin has been based on the leading author.

\begin{figure}[h]
%	\begin{addmargin}{\dimexpr -\oddsidemargin-0.3in\relax}
	\includegraphics[width=\textwidth]{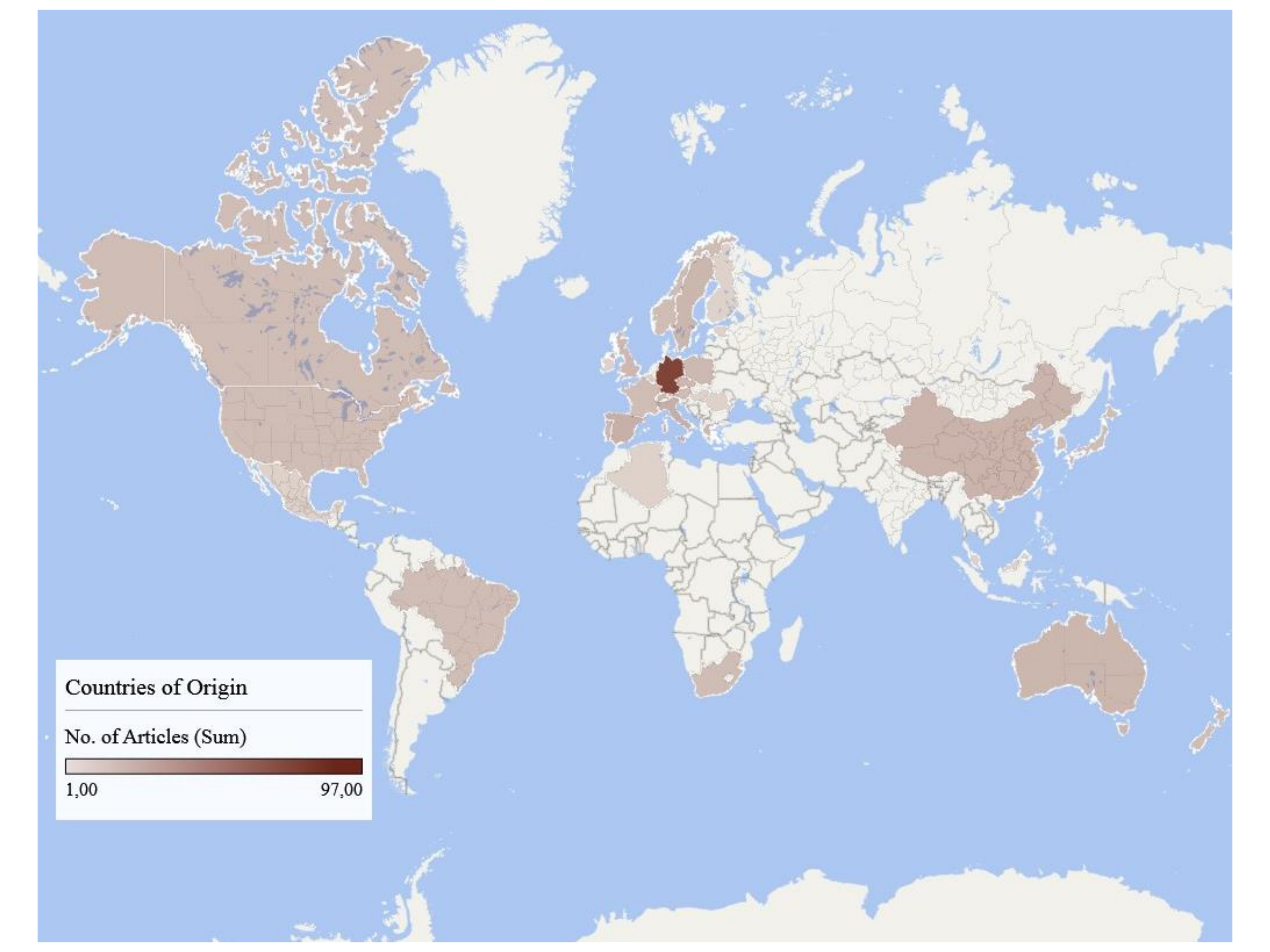}
%	\end{addmargin}
	\caption{Countries of Origin of Sample Articles}
	\label{figure:countries_continents_of_origin}
\end{figure}

Overall, publications from 32 different countries are included in the final sample. Ten of those countries appear exactly once, the same number of countries is represented twice among the sample. Another six countries, all from Europe, appear four times within the sample. Finally, five publications stem from Australia and Sweden, respectively, seven articles originated in China, ten in Spain, eleven in Italy, and 97 in Germany, which seems less striking when considering the overall origination of the \ac{I4.0} concept. 

Summarizing on a broader level, 1.1\% of publications included in the final sample stem from South America, 1.6\% from Africa, 2.7\% from North America, 3.7\% from Australia and Oceania, 6.9\% from Asia, and the vast majority of 84.1\% from a European country. The distributions of countries and continents of origin are illustrated in the form of corresponding heatmaps in Figure \ref{figure:countries_continents_of_origin}.

\subsection{Analysis and Structuring of Literature Contents in the Final Sample}
\label{sec:sample_analysis}

Based on the collected and filtered final sample of relevant scientific literature, fundamental data on the articles' contents has been collected and documented for the development of a taxonomy of \ac{HMI} in \ac{I4.0} as well as for all subsequent analyses in this paper. For this purpose, every paper of the final sample has been read and analyzed thoroughly in order to identify the entirety of aspects of \ac{HMI} in \ac{I4.0} addressed by the respective article. 

To be more precise, a Microsoft Excel spreadsheet has been created serving as a tool for detailed documentation. A row has been dedicated for every article while each column, besides the one for the respective year of publication, represents a specific attribute identified as relevant to the development of a taxonomy of \ac{HMI} in \ac{I4.0} during perusal of sample papers. In the course of analyzing an increasing number of papers, this collection of attributes had grown to reach an interim number of 154. Accordingly, the contents of all 189 sample articles have been analyzed, documenting for every article and attribute, in the respective cell of the spreadsheet, whether the attribute is addressed in the respective paper.

In order to derive a meaningful taxonomy of \ac{HMI} in \ac{I4.0} from this massive and largely unstructured collection of data, the identified attributes have been clustered according to superordinate categories of the topic which they represent instances of. These categories, again, have been clustered according to the topic's focal points they address and then organized by arranging them in a hierarchy. 

In doing this, the overall concept of \ac{HMI} in \ac{I4.0} has been structured and systematically mapped in a taxonomy classifying the topic according to categories arranged in a three-level hierarchy. The structure of the hierarchy implies a detailing and further partitioning of elements from first-level categories down to third-level categories. Below the third level, the initial attributes are grouped according to the respective category which they represent a specification of. 

For the final taxonomy and the following analyses of the paper, similar attributes have been merged and dispensable categories and attributes not essential to a comprehensive but, at the same time, focused description of the topic have been removed. Therefore, the final taxonomy and its three-level hierarchy of thematic categories are based on a foundation of 113 attributes for which coverage in each of the 189 sample articles has been documented thoroughly in the spreadsheet.

\subsection{Analysis of the Data Collected for Content Documentation}
\label{sec:data_analysis}

Besides the initial development and derivation of the actual taxonomy, the detailed documentation of the article contents has served another purpose. In particular, the collected data set enables a quantitative analysis of the distribution and patterns of research interest among different thematic aspects complementing the qualitative structuring of the topic into a taxonomy of \ac{HMI} in \ac{I4.0}. The following subsections outline the methodology and steps implemented for this purpose which are an analysis of the relative frequencies of different taxonomy elements among the final sample as well as the application of machine learning algorithms for the purpose of data mining. 

\subsubsection{Analysis of the Relative Frequencies of Individual Taxonomy Elements}
\label{subsec:analysis_relative_frequencies}

In order to reveal potential disparities in the coverage of individual taxonomy elements among the entirety of sample articles, the total number of papers addressing each respective attribute of the taxonomy has been determined. Dividing this number by the overall size of the sample (189) has yielded the relative frequency of an attribute, stating the percentage share of articles discussing the concerned element. Knowing the relative frequencies allows for a detection of those instances representing a focal point of research on the topic of \ac{HMI} in \ac{I4.0}. 

However, general research streams might not concentrate only on specific bottom-level attributes but, instead, focus on larger subsections of the taxonomy, represented by higher-level categories. In particular, if different authors consider varying attributes or manifestations of the same category, they still address the same higher-level aspect of the overall topic. Therefore, the Microsoft Excel documentation of attribute coverage has been extended to reflect also the prevalence and relative frequencies of higher-level categories among the final sample. For this purpose, a specific category has been marked as addressed by a respective article if the same applies to at least one of the attributes belonging to this category.

\subsubsection{Data Mining on the Documented Literature Contents}
\label{subsec:analysis_data_mining}

As a final step in the analysis of the taxonomy, data mining techniques have been applied to the underlying data reflecting the contents of the sample articles. The aim of mining the documented article contents is to uncover underlying patterns of research streams and related sub-aspects of the topic as potentially indicated by associations and interrelations present among the content-related sample data. For this purpose, a two-fold approach has been applied in order to scrutinize the collected data for both implicit association rules and classification rules to be derived from the data.

Regarding implementation, a set of machine learning algorithms has performed the data mining process. For that purpose, the open-source \ac{Weka} workbench developed at University of Waikato, New Zealand has been deployed which provides a rich set of readily available machine learning algorithms \cite{Witten2013}. For the analyses in the course of this paper, three of those algorithms have been deployed. Those are the Apriori algorithm searching for the best association rules according to specifiable parameters as well as the classifiers ZeroR and JRip.

Initially, following the \ac{CRISP-DM} process model, the aims targeted at in applying data mining techniques have been determined by formulating the research questions RQ3.1 and RQ3.2 (business understanding). 

Based on the collected set of data available and the research questions pursued, the decision has been made to reduce the data set reflecting sample literature contents to a level of specifying prevalence of only the third-level categories of the taxonomy. Thus, individual attributes and categories of the first and second level as well as data on their coverage in sample articles have been removed from the data set (data preparation). 
%A representation of the prepared data set for data mining purposes is provided in Appendix \ref{sec:apx_data_set_weka}. 

Determined in the data understanding phase, the reason for studying data on third-level categories instead of attributes of the taxonomy is the lower number of categories compared to attributes. Thus, chances are higher for the algorithms to identify meaningful associations and relations in the data due to a lower ratio of attributes\footnote{\texttt{Attributes} in the context of \ac{Weka} applications refer to the properties of the data set used as input for the machine learning algorithms, not to the bottom-level elements of the taxonomy.} to sample instances in the data set \cite{Witten2013b}. First- and second-level categories, on the other hand, are discarded in order to avoid auto-correlation among sample data, given that coverage of a lower-level category implies coverage of its superordinate categories.

The first algorithm applied on the reduced data set in the modeling phase is the Apriori algorithm to find association rules within the data \cite{Witten2014}. This means that item sets, i.e. attribute values or combinations of values of different attributes, are searched which, when observed within a sample instance, appear together with a specific value of another attribute \cite{Witten2014}.

Finally, the \ac{Weka} workbench has also been applied in order to identify models classifying instances regarding the value of a specific attribute in the data set, called the \texttt{class} \cite{Witten2013}. The aim for the algorithm deployed is to determine a set of e.g. rules or a function which enables predicting the value of the class if the set of values of the other attributes is known for a specific instance \cite{Witten2013}. For this purpose, the algorithm requires a data set as input, called the \texttt{training set}, used for learning and identifying the classification model. The aim is to determine a model classifying any conceivable instance from the real world, i.e. of the overall population, as accurately as possible. The accuracy of the model derived from the training set, i.e. from an assumed random sample of the population, should be evaluated on another independent sample, the \texttt{test set} \cite{Witten2013a}. 

Weka offers stratified cross-validation as a testing option for classification models which is used with ten folds in this study as part of the evaluation phase of \ac{CRISP-DM}. In ten-fold stratified cross-validation, the group of instances in the overall data set is split into ten equal parts where, in each fold, the values of the class attribute are distributed according to their distribution in the overall data set \cite{Witten2013a}. Then the algorithm performs ten rounds of building a classification model based on a training set comprising nine of the ten folds and each time tests the model's classification accuracy on the tenth fold whereby each fold serves as the test set once. In an eleventh round, the algorithm uses the entire data set as training set to develop the final classification model whose accuracy is estimated to be the mean of the tested accuracies of the previous ten rounds \cite{Witten2013a}.

For the purpose of this study, the used classification algorithm has been run ten times for every class attribute using a different seed value for the randomization procedure inherent to the algorithm in each round. The mean of the ten accuracies, each one estimated using ten-fold stratified cross-validation respectively, is supposed to represent a more reliable estimate of the true accuracy of classification models produced by the algorithm for the respective class \cite{Witten2013a}.

From the multitude of available classification algorithms in the \ac{Weka} workbench, JRip has been chosen to identify a model for each attribute. However, the classification model produced by JRip has not been accepted in case its estimated accuracy turned out lower than the baseline accuracy of the classification model delivered by the ZeroR algorithm (evaluation phase). The ZeroR model neglects any attribute value in predicting an instance's class value because ZeroR only delivers one simple unconditional classification rule stating that the class value of any instance is predicted to be the specific class value appearing most frequently among all instances of the training set \cite{Witten2013a}.

As another means of verifying the meaningfulness of the models accepted based on a comparison to the baseline accuracy, \ac{ROC} area values of those models have been examined which should be greater than 0.5 (cf.~\cite{Powers2011}). A more detailed rationale for applying JRip and considering ZeroR and \ac{ROC} area values for evaluation purposes is provided in Section \ref{subsec:rq3_2}.

\section{Results RQ1: A Taxonomy of Human-Machine Interaction in Industry 4.0}
\label{chap:results_taxonomy}

After the detailed description of the methodology applied in this study, the following section presents the initial principal outcome and result of the implemented analyses. In particular, this means that the first main research question motivating this paper, i.e. \ac{RQ1}, is supposed to be answered by providing a structuring of the overall topic of \ac{HMI} in \ac{I4.0} in terms of a comprehensive taxonomy.

\begin{figure}[h]
	\begin{center}
		\includegraphics[width=10.5cm]{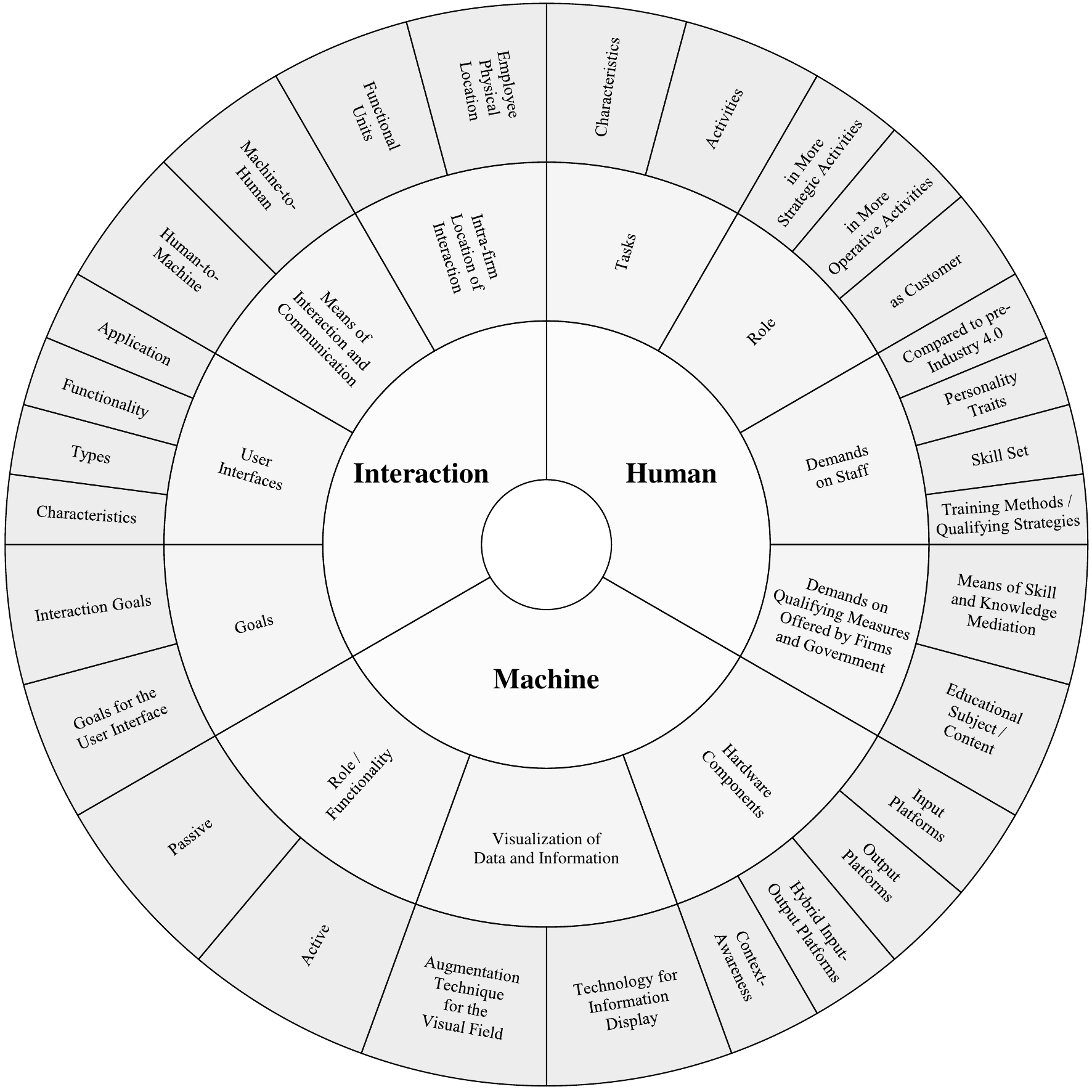}
		\caption{A Taxonomy of \acs{HMI} in \acs{I4.0}}
		\label{figure:taxonomy_total} 
	\end{center}
\end{figure}

As stated in subsection \ref{sec:research_questions}, \ac{RQ1} asks what the current state of the art in \ac{HMI} research in the area of \ac{I4.0} is. In order to holistically capture the concept of \ac{HMI} in \ac{I4.0} as discussed in scientific literature, its various facets and sub-aspects have been captured, systematically structured, and described in terms of a comprehensive taxonomy. This model, illustrated in Figure \ref{figure:taxonomy_total}, reflects the focal points and analytic approaches towards the topic deployed in related research.

As can be deduced from Figure \ref{figure:taxonomy_total}, the taxonomy is subdivided into three top-level categories corresponding to the main features of the topic represented by \texttt{Human}, \texttt{Machine}, and \texttt{Interaction}. The rationale behind this classification is to highlight the distinction between the two major types of intrinsically different agents in \ac{I4.0} scenarios, i.e. the human operator and the machine entity. Still, in the course of \ac{HMI} processes, these two types of actors are supposed to interact and collaborate purposefully and successfully as an integral part of \ac{I4.0} activities. This potentially complex and sophisticated interaction implicates a variety of different facets to analyze and take into account if \ac{HMI} in an \ac{I4.0} environment is supposed to function appropriately which is why the third category besides the human and machine entity is dedicated towards their interaction.

Shifting the focus to a broader perspective of examining the overall taxonomy, Figure \ref{figure:taxonomy_total} reveals that, on the second level of hierarchy, both the human and interaction category comprise four sub-categories, respectively, while the machine part is detailed into three spheres on the intermediate level. Finally, on the lowest level of categories constituting the taxonomy, the human sector comprises eleven distinct elements, the interaction part entails ten subcategories, and the machine subsection adds another eight detailed categories to the taxonomy. 

In the following, this section describes each of the three top-level categories and their respective lower-level elements in more detail, including overviews over the detailed attributes representing the bottom level of the hierarchy. 

\subsection{The \texttt{Human} Dimension}
\label{sec:tax_human}

\begin{figure}[htb]
		\includegraphics[width=0.5\textwidth]{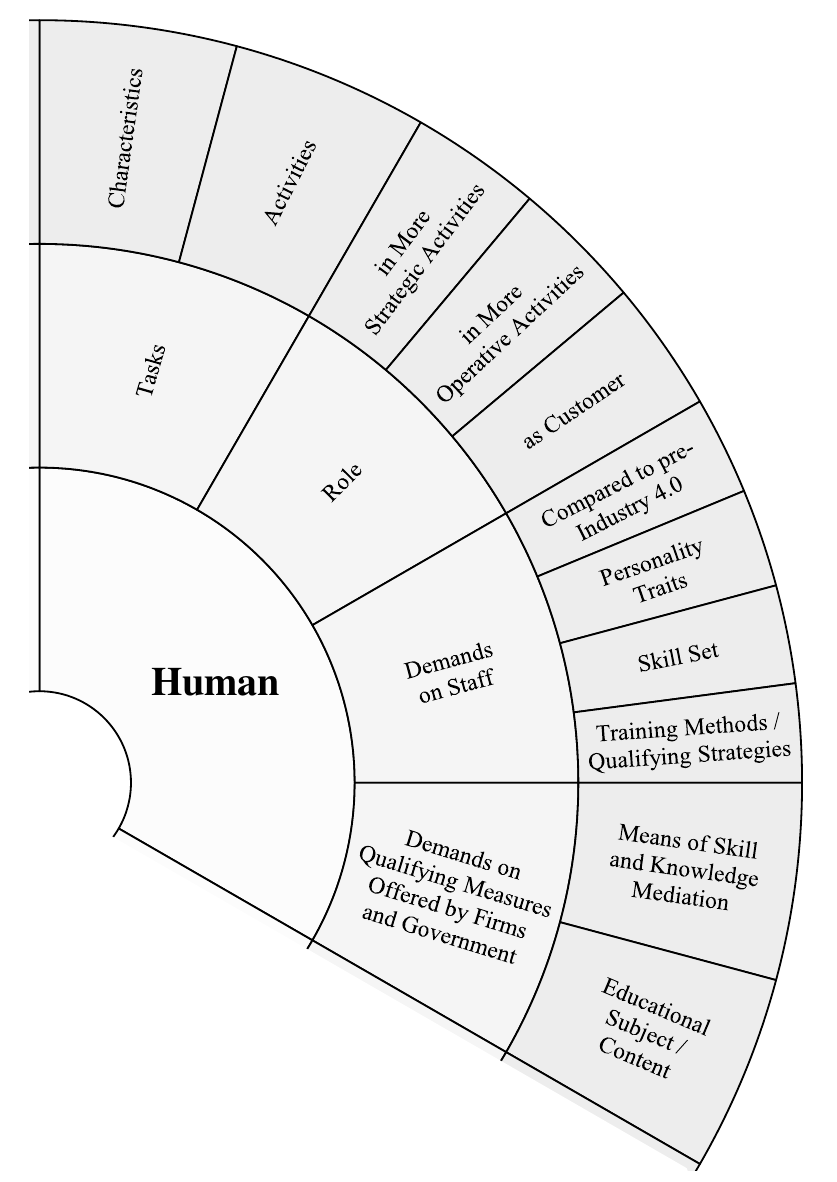}
	\caption{A Taxonomy of \acs{HMI} in \acs{I4.0} - \texttt{Human} subsection}
	\label{figure:taxonomy_human}
\end{figure}

Starting with the \texttt{Human} subsection, the structure of the taxonomy shall be described down to the lowest level of elements, i.e. the attributes. For that reason, Figure \ref{figure:taxonomy_human} displays the corresponding subsection of the overall taxonomy, complemented by a table specifying the respective attributes of the bottom-level categories.

Considering the human entity in processes of \ac{HMI} in \ac{I4.0}, four major aspects can be identified and distinguished, namely the tasks to be performed by human operators in future manufacturing activities, the role humans are supposed to assume in complex \ac{I4.0} production systems, the demands placed on staff in order to be able to fulfill those designated tasks and roles, and the requirements regarding the available educational and qualifying measures to be provided by governmental and corporate entities in order to provide for the future \ac{I4.0} workforce.

\begin{table}[]
    \centering
    \begin{tabular}{L{2.3cm} p{6cm}}
			{\textbf {\footnotesize 3rd-Level Category}} & {\textbf {\footnotesize Attributes}} \\ \hline
			\hline
			{\textbf {\tiny Characteristics}} & {\tiny \{new; manifold/diverse; fast-changing; short-term/ hard-to-plan; process- instead of technology-oriented\}} \\
			\rule{0pt}{13pt}{\textbf {\tiny Activities}} & {\tiny \{planning/specification/engineering; monitoring; verification; interpretation of prepared data \& information; machine operating; manual assembly; troubleshooting/intervention in case of failure\}} \\ \hline
			{\textbf {\tiny in More Strategic Activities}} & {\tiny \{decision-maker; coordinator of value creation; producer/source of knowledge\}} \\
			\rule{0pt}{13pt}{\textbf {\tiny in More Operative Activities}} & {\tiny \{flexible problem-solver; more responsibility; controlling instance and supervisor of ongoing activities\}} \\
			\rule{0pt}{13pt}{\textbf {\tiny as Customer}} & {\tiny \{participation in product design\}} \\ \hline
			{\textbf {\tiny Compared to pre-Industry 4.0}} & {\tiny \{new demands; extended demands; further training required\}} \\
			\rule{0pt}{13pt}{\textbf {\tiny Personality Traits}} & {\tiny \{flexibility; responsibility\}} \\
			\rule{0pt}{13pt}{\textbf {\tiny Skill Set}} & {\tiny \{abstraction/comprehensive process thinking; inter-disciplinary skills; social skills; media skills; quantitative/IT skills; technological skills\}} \\
			\rule{0pt}{13pt}{\textbf {\tiny Training Methods / Qualifying Strategies}} & {\tiny \{lifelong learning/continuous professional development; inter-employee knowledge transfer; training on the job/work-integrated learning\}} \\ \hline
			{\textbf {\tiny Means of Skill and Knowledge Mediation}} & {\tiny \{new training forms/educational system; learning factories; personalized training; round-based training concept; virtual training\}} \\
			\rule{0pt}{13pt}{\textbf {\tiny Educational Subject / Content}} & {\tiny \{training on new technologies; \acs{MINT} education\}} \\ \hline
		\end{tabular}
    \caption{Dimensions of the \texttt{Human} subsection}
    \label{tab:human}
\end{table}

\textbf{Tasks.} The tasks of human operators in \ac{I4.0} operations, in turn, can be described in terms of their characteristics and the actual types of activities they encompass. According to insights from relevant scientific literature included in the final sample, human tasks in \ac{I4.0} will be diverse and novel and continue to evolve quickly (e.g. \cite{Fantini2016,Kagermann2015}). While some authors expect human tasks to be highly process-oriented (e.g. \cite{Wagner2017}), human operators will also be expected to perform those spontaneous activities that are difficult to foresee and plan in advance (e.g. \cite{Stock2016}).

Accordingly, an important type of human tasks in \ac{I4.0} operations will be the intervention in case of failure (e.g. \cite{Prinz2016}). Besides, an important scope of human activities will be the mental work of planning and engineering, thus specifying processes and production plans whose implementation and consequential output human operators then monitor and verify, respectively (e.g. \cite{Gorecky2014}). On the other hand, \ac{I4.0} activities might still imply more hands-on tasks for human operators, in particular manual assembly and machine operating (e.g. \cite{Hao2017}). What combines most of those activities in \ac{I4.0} is that the vast majority might, to some degree, entail the interpretation of data and information provided for the execution of a specific task (e.g. \cite{Wittenberg2016}).

\textbf{Roles.} In performing the tasks described, human operators will assume specific roles within \ac{I4.0} processes, especially with regard to a differentiation from the role of machine entities in \ac{HMI} procedures. With a more strategic outlook, employees are supposed to act as decision-makers and coordinators concerning the alignment of value creation processes (e.g. \cite{Stock2016}). Besides, human agents are seen as a source and potential producers of knowledge in the organization (e.g. \cite{Posada2015}).

From a more operative perspective, human operators are expected to assume the role of a flexible problem-solver and of a controlling instance and supervisor of ongoing activities. Altogether, this implies an increase in the degree of responsibility employees throughout the organizational structure, down to shop floor operators, are supposed and entitled to assume (e.g \cite{Gorecky2014}). 

Finally, a couple of authors also shift the focus on the human role away from solely considering employees and towards the role as a customer in an \ac{I4.0} environment where innovative digital and online solutions enable the customer to actively participate in product design or configuration (e.g. \cite{Wang2017}).

\textbf{Demands on Staff.} Coming back to the intra-firm perspective on employees, the demands placed on staff in comparison to industrial activities prior to \ac{I4.0} inception are considered to be partly new and extended which, as a consequence, requires employees to undergo further training (e.g. \cite{Saggiomo2016}). 

With regard to the expected personality traits of employees, working in an \ac{I4.0} environment will require both a certain degree of personal flexibility and a necessary sense of responsibility (e.g. \cite{Benesova2017}). 

Looking more precisely at the actual skills employees are expected to provide, these can be both more generalist or more specialized. This means that the required skill set might comprise the capability for abstraction and comprehensive process thinking, social skills, and inter-disciplinary skills (e.g. \cite{Hirsch-Kreinsen2014}), as well as quantitative and IT, media, and technological skills (e.g. \cite{Benesova2017}).

Considering the extent and broadness of demands on staff, employees will need to be willing to undergo qualifying measures, likely in the form of lifelong learning during the entire professional career (e.g. \cite{Zhou2015}). Besides, learning on the job and from colleagues in the course of inter-employee knowledge transfer represent vital opportunities for staff to acquire crucial skills and knowledge (e.g. \cite{Pfeiffer2016}).

\textbf{Demands on Qualifying Measures.} The last interim-level category picks up on the aspect of qualifying the future \ac{I4.0} workforce, specifying demands placed on governments and firms concerning educational and training opportunities to be offered. Regarding the actual educational content offered, sample literature mainly concentrates on training on new technologies and education in \ac{MINT} subjects (e.g. \cite{Kagermann2015}).

Besides, the taxonomy differentiates among specific means of knowledge and skill mediation addressed among the sample articles. While, in general, new training forms and an adjustment of the educational system are requested (e.g. \cite{Thiede2016}), learning factories and personalized training are the most often-cited specific means of effectively imparting necessary knowledge to trainees (e.g. \cite{Gronau2017}). In addition, forms of computer-aided virtual training and a round-based training concept, e.g. incorporating gamification elements, are considered to be potentially appropriate and useful (e.g. \cite{Gorecky2015}).

\subsection{The \texttt{Machine} Dimension}
\label{sec:tax_machine}

\begin{figure}[h]
	\centering
	\includegraphics[width=10cm]{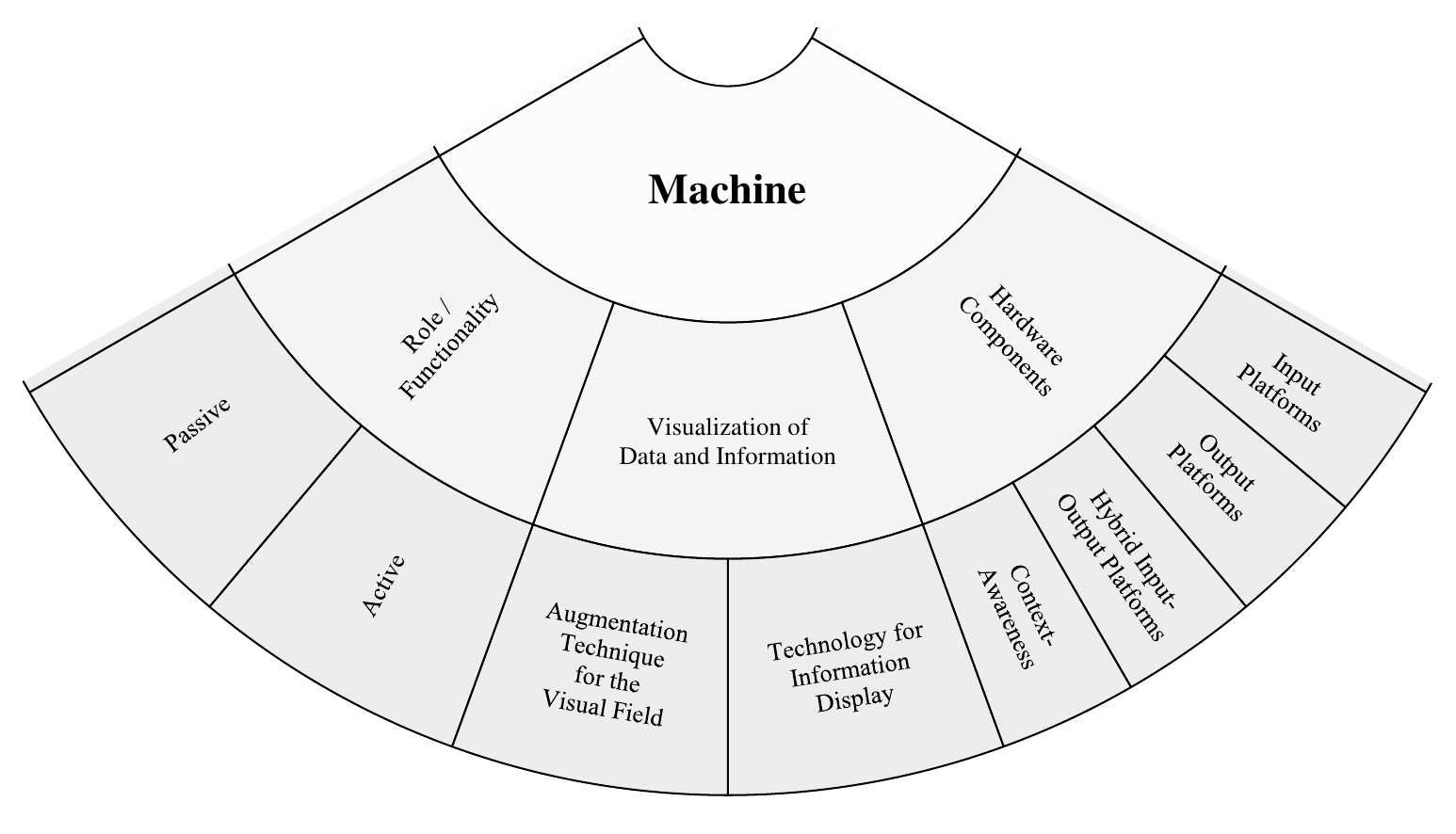}
	\caption{A Taxonomy of \acs{HMI} in \acs{I4.0} - \texttt{Machine} subsection}
	\label{figure:taxonomy_machine}
\end{figure}

As indicated above, the \texttt{Machine} subsection of the taxonomy is split into only three interim-level categories, which describe and define the role and functionality machine entities assume and provide in \ac{HMI} processes as opposed to the role and the tasks of human actors, the type and means of visualization of data and information, and the hardware components deployed for those purposes. Again, the corresponding part of the overall taxonomy, complemented by a table listing the respective attributes of the third-level categories, is illustrated in Figure \ref{figure:taxonomy_machine}.

\begin{table}[]
    \centering
	\begin{tabular}{L{3.5cm} p{10cm}}
		{\textbf {\footnotesize 3rd-Level Category}} & {\textbf {\footnotesize Attributes}} \\ \hline
		\hline
		{\textbf {\tiny Passive}} & {\tiny \{collection, aggregation, preparation, and visualization of data \& information; monitoring of manual workflow; (self-)diagnostics\}} \\
		\rule{0pt}{13pt}{\textbf {\tiny Active}} & {\tiny \{implementation; support and assistance for the human; collaboration with humans on joint tasks\}} \\ \hline
		{\textbf {\tiny Augmentation Technique for the Visual Field}} & {\tiny \{adding realistic virtual objects; modifying the user's view; adding graphics from traditional illustration (distinct from reality)\}} \\
		\rule{0pt}{13pt}{\textbf {\tiny Technology for Information Display}} & {\tiny \{video see-through; optical see-through; (in-situ) projection\}} \\ \hline
		{\textbf {\tiny Context-Awareness}} & {\tiny \{sensors; positioning systems; RFID chips\}} \\
		\rule{0pt}{13pt}{\textbf {\tiny Hybrid Input-Output Platforms}} & {\tiny \{hand-held (e.g. tablet, smartphone)\}} \\
		\rule{0pt}{13pt}{\textbf {\tiny Output Platforms}} & {\tiny \{stationary screen/monitor; \ac{HMD} (stereoscopic); \ac{HMD} (monoscopic); \ac{HMD} (no further details)\}} \\
		\rule{0pt}{13pt}{\textbf {\tiny Input Platforms}} & {\tiny \{manual controllers; other mechanical control devices (e.g. keyboard, foot pedal); camera (image capturing); wearable sensors (motion capturing)\}} \\ \hline
	\end{tabular}
    \caption{Dimensions of the \texttt{Machine} subsection}
    \label{tab:machine}
\end{table}

\textbf{Role/Functionality.} As can be inferred from Figure \ref{figure:taxonomy_machine}, the machine entity can assume both a passive and an active role in \ac{HMI} procedures under \ac{I4.0}, providing functionality either rather passively or rather proactively, while an exact definition of the boundaries depends on the individual case.

Assuming a passive role, the machine entity can provide a functionality of monitoring manual workflows (e.g. \cite{Weyer2015}) or diagnosing roots of emerging problems, both machine-internal and external (e.g. \cite{Wittenberg2016,Romero2016a}). A particularly crucial passive functionality is the continuous collection, aggregation, preparation, and visualization of data and information (e.g. \cite{Gorecky2014}), e.g. in order to facilitate human operators' fulfillment of the role of a strategic decision-maker.

Providing more active functionality, the automated system actively supports and assists the human actor, also comprising active physical support as a means of compensating e.g. for decreasing human capabilities with rising age of employees (e.g. \cite{Saggiomo2016}). Besides acting as a mere assistant for human operators, machine entities can assume the role of an active and equal collaborator of humans on joint activities, sharing and dividing tasks according to the respective strengths (e.g. \cite{Wang2015}). As another active functionality complementary to the tasks of the human operator specified in subsection \ref{sec:tax_human}, the machine entity can implement the production strategies and plans specified by a human (e.g. \cite{Gorecky2014}).

\textbf{Information Visualization.} Concerning visualization of data and information, scientific literature differentiates among various techniques to augment the perceived view of the recipient of information while there are different technologies available to display this information to the human eye.

For the former, literature differentiates among three techniques to modify and enhance the visual perception of a human user which are examined and deployed in various combinations in different articles. The visual field can be enhanced by adding realistic \ac{3D} virtual objects to the real-world view or by adding common graphical elements from traditional illustration, e.g. arrows, which are clearly distinct from the underlying view of reality. Besides, the general view of the environment can be modified by changing contrast or color saturation (e.g. \cite{Paelke2014}). 

As a means of implementing those visual augmentations, different technologies are available to display information to the human eye. An intuitive solution is the use of video see-through technology displaying digitally enhanced camera footage of the real world on a screen (e.g. \cite{Pirvu2016}). As an alternative technology, optical see-through is available, allowing the user to retain a view of the actual environment through a transparent medium like glass and displaying only the added virtual objects digitally (e.g. \cite{Fraga-Lamas2018}). Hence, in contrast to video see-through solutions, the user's view of the real world will remain intact in case of a shutdown of the digital display \cite{Syberfeldt2015}. This advantage also applies to the third information display technology covered in sample articles which is in-situ projection deploying physical objects from the environment as a screen surface upon which a projector projects digital information and images (e.g. \cite{Funk2016}).

\textbf{Hardware Components.} In order to complement the description of the machine entity in \ac{HMI} processes under \ac{I4.0}, the taxonomy includes a specification of the hardware components employed in order to implement the machine's functionality and information visualization. In this context, articles included in the final sample discuss different ways of achieving context-awareness of the automated system and different hardware platforms for user input and visual output as well as hybrid input-output platforms.

As a means of achieving context-awareness of \acp{CPS}, a multitude of sensors like power meters, accelerometers, thermometers, or others \cite{Liu2017} as well as \ac{RFID} chips and positioning systems like \ac{GPS} and indoor solutions (e.g. \cite{Gorecky2014}) are available. 

Besides, an important aspect of \ac{HMI} process design in \ac{I4.0} is the choice of respective input and output platforms. In this regard, common hand-held devices from the consumer market, i.e. mainly tablets and smartphones, represent a hybrid solution potentially suitable for industrial applications as well (e.g. \cite{Masoni2017}).

Regarding dedicated output platforms used for visualization purposes, different studies address either stationary screens and monitors (e.g. \cite{Loch2016}) or different types of \acp{HMD} like smart glasses although authors only rarely define whether the respective study focuses on stereoscopic or monoscopic \acp{HMD} (e.g. \cite{Jost2017}).

Finally, designated input platforms serving as a means for human user input can be represented by mechanical devices like manual controllers (e.g. \cite{Linn2017}) or other control devices like keyboards and foot pedals (e.g. \cite{Funk2017}). Besides, user input can be registered by cameras in the form of image capturing of the user and by wearable sensors directly capturing user motion (e.g. \cite{Khalid2016}).

\subsection{The \texttt{Interaction} Dimension}
\label{sec:tax_interaction}

\begin{figure}[h]
		\includegraphics[width=5cm]{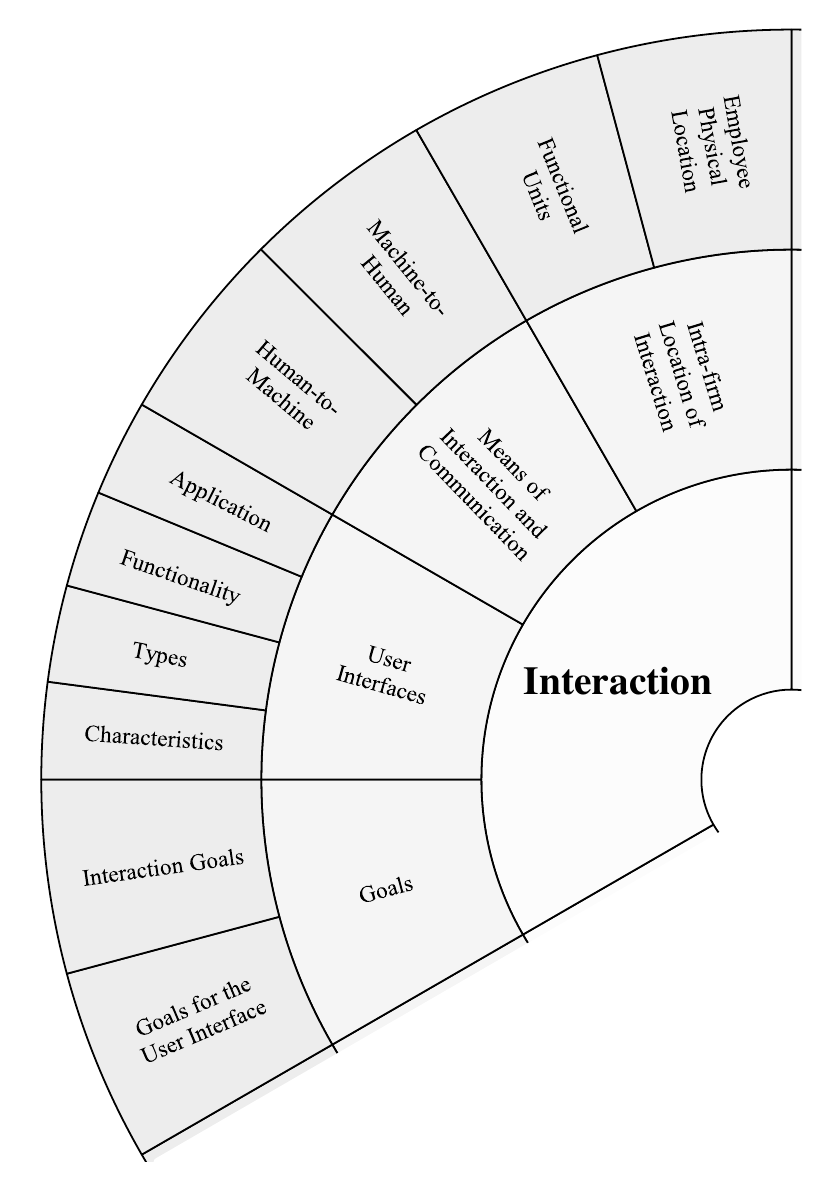}
	\caption{A Taxonomy of \acs{HMI} in \acs{I4.0} - \texttt{Interaction} subsection}
	\label{figure:taxonomy_interaction}
\end{figure}

The subsection of the taxonomy describing the character of interaction between human and machine entity in \ac{I4.0} is split into four categories defining the location and the means of interaction, the \acp{UI} employed, and the goals of interaction. Figure \ref{figure:taxonomy_interaction} provides a comprehensive overview over this \texttt{Interaction} part including a detailed enumeration of the attributes specifying the third-level categories.

\begin{table}[]
    \centering
		\begin{tabular}{L{2.5cm} p{7cm}}
			{\textbf {\footnotesize 3rd-Level Category}} & {\textbf {\footnotesize Attributes}} \\ \hline
			\hline
			{\textbf {\tiny Employee Physical Location}} & {\tiny \{fixed/workplace-bound; arbitrary\}} \\
			\rule{0pt}{13pt}{\textbf {\tiny Functional Units}} & {\tiny \{production/manufacturing; (intra-)logistics\}} \\ \hline
			{\textbf {\tiny Machine-to-Human}} & {\tiny \{messages to human's device; haptics\}} \\
			\rule{0pt}{13pt}{\textbf {\tiny Human-to-Machine}} & {\tiny \{multi-touch/touch screen; voice control; gesture recognition/mimics; eye gaze\}} \\ \hline
			{\textbf {\tiny Application}} & {\tiny \{planning, design, and simulation; training; process visualization/production monitoring \& quality control; maintenance; (remote) robot control; instructions for the human; navigation; value-added services for providers \& clients\}} \\
			\rule{0pt}{13pt}{\textbf {\tiny Functionality}} & {\tiny \{context-/problem-specific reduction of complexity; tracking of component positions; tracking of human position/motion; adding of knowledge components; providing individual feedback; \acs{3D} interactive customer tool\}} \\
			\rule{0pt}{13pt}{\textbf {\tiny Types}} & {\tiny \{\acs{VR}; \acs{AR}; traditional \acs{GUI} (\acs{WIMP}); \acs{NUI} (post-\acs{WIMP})\}} \\
			\rule{0pt}{13pt}{\textbf {\tiny Characteristics}} & {\tiny \{intelligent; web-based; mobile; context- \& user group-sensitive\}} \\ \hline
			{\textbf {\tiny Interaction Goals}} & {\tiny \{augmentation of human capabilities; boost of human productivity \& quality of work; reduction of human errors; improvement of human decision quality; improvement in human intervention into automated processes; increase of manufacturing flexibility; customization; improvement of working conditions; reduction of professional exclusion; increase of training efficiency; increase of human engagement\}} \\
			\rule{0pt}{13pt}{\textbf {\tiny Goals for the User Interface}} & {\tiny \{usability; transparency of system behavior; optimal \& dynamic task allocation between human \& machine; adaptivity \& self-learning; adaption to current situation at run-time; identification \& evaluation of current work procedure\}} \\ \hline
		\end{tabular}
    \caption{Dimensions of the \texttt{Interaction} subsection}
    \label{tab:interaction}
\end{table}

\textbf{Location of Interaction.} The \texttt{Interaction} part of the taxonomy analyzes the intra-firm location of \ac{HMI} with respect to the organizational structure, i.e. functional units, and the physical location of interaction. The latter can be either fixed to a specific workplace where \ac{HMI} occurs, or modern technologies featuring ubiquitous connectivity and information availability enable arbitrary interaction possibilities for the human operator (e.g. \cite{Gorecky2014}).

Apart from that, considering the topic, it seems little surprising that a majority of sample articles analyzes \ac{HMI} in a production or manufacturing environment. However, the implications of \ac{I4.0} for \ac{HMI} processes are also scrutinized regarding logistics operations, both internal and beyond company borders (e.g. \cite{Barreto2017}).

\textbf{Means of Interaction.} A decisive aspect when designing \ac{HMI} modules for modern \ac{I4.0} production systems is obviously the actual means of interaction or, more precisely, of communicating bidirectionally. In this context, the taxonomy distinguishes between \ac{H2M} and \ac{M2H} communication. A way for the system to communicate to the user, in that sense, is to convey messages to operators via their end devices (e.g. \cite{Kolberg2015}) or to control their attention physically via haptic impulses and feedback (e.g. \cite{Posada2015}). 

More intuitive ways of communication and interaction have also been established for industrial \ac{H2M} communication in the form of touch interfaces (\cite{Flatt2015}), natural language interfaces, i.e. voice control, gesture control including machines mimicking human motion, or even interaction via human gaze, implemented by the machine tracking and interpreting the user's line of sight (e.g. \cite{Pirvu2016}). 

\textbf{User Interfaces.} A lot of attention in research related to \ac{HMI} in \ac{I4.0} is spent on \acp{UI} which the taxonomy describes in terms of their characteristics, types, functionality, and application scenarios, the latter being wide and diverse. \acp{UI} are deployed for \ac{HMI} processes in planning, design, and scenario simulation during product development, in training, monitoring and quality control, as well as in maintenance activities (e.g. \cite{Romero2016a}). More precisely, \acp{UI} are applied to convey instructions to operators, to provide navigation services, i.e. guidance (e.g. \cite{Blanco-Novoa2018}), or to enable operators to control industrial robots, even remotely (e.g. \cite{Wang2015}). Besides, advanced \acp{UI} can be used to provide value-added services to providers and especially clients, e.g. by supplying digital manuals for installation and/or maintenance of delivered products (e.g. \cite{Posada2015}). 

In order to be suitable for these application scenarios, \acp{UI} provide various functionalities. As a means to avoid overwhelming the user with an abundance of information and selection options, \acp{UI} should implement context- and problem-specific reduction of complexity and information provision (e.g. \cite{Gorecky2014}). In order to enable navigation and accurate instructions, tracking of relevant components in the environment and tracking of the actual user are required (e.g. \cite{Pirvu2016}). Considering the size of regular industrial organizations and the number of human operators, a \ac{UI} should allow for adding of knowledge components by the user in order to make them available and accessible for other users via the \ac{UI} (e.g. \cite{Flatt2015}). At the same time, providing worker-individual feedback is important in many applications like accurate assembly instructions (e.g. \cite{Funk2016}). Finally, in reference to the human role of a customer involved in product design, a potential \ac{UI} offered online to customers is supposed to implement a \ac{3D} interactive tool for individual product configuration (e.g. \cite{Zawadzki2016}). 

Regarding possible types of \acp{UI}, a typical \ac{UI} still occasionally employed or studied for \ac{I4.0} purposes is the traditional \ac{GUI} implementing \ac{WIMP} properties (e.g. \cite{Thiede2016}). However, more innovative and intuitive forms of \acp{UI} are on the rise and intensely studied in various sample articles. In general, \acp{UI} of a post-\ac{WIMP} type, so-called \acp{NUI}, are addressed in order to facilitate and foster \ac{HMI} processes in \ac{I4.0} (e.g. \cite{Paelke2014}). In particular, \acp{UI} implementing \ac{VR} or especially \ac{AR} have received significant attention among \ac{I4.0}-related research (e.g. \cite{Pfeffer2015}).

Irrespective of the specific type implemented, \acp{UI} in an \ac{I4.0} context are mainly characterized by one or several of the following properties: While generally being described as intelligent in a number of articles, \acp{UI} might, in particular, provide context- and user group-sensitivity, implying a capability to adapt features like layout, options, and information offered to the current usage situation at hand (e.g. \cite{Gorecky2014}). Furthermore, depending on the application scenario, various studies discuss the use of mobile and potentially web-based \acp{UI} (e.g. \cite{Moreno2017}).

\textbf{Goals.} In conclusion of both the description of the \texttt{Interaction} part and of the structure of the overall taxonomy, the \texttt{Goals} subsection consists of goals for the \ac{UI} and goals to be achieved by implementing successful \ac{HMI} in general. 

Concerning the latter, on an individual employee level, human capabilities are supposed to be enhanced through the interplay with machines in their assistive role described in subsection \ref{sec:tax_machine} which shall, in turn, lead to an enhancement of human productivity and quality of work and a reduction in the number of operators' errors (e.g. \cite{Romero2016}). Besides, context-specific provision of data and information in \ac{HMI} aims at improving human decision quality (e.g. \cite{Venkatapathy2017}). 

From an overall organizational perspective, \ac{HMI} aims for a general increase in productivity by means of improved human intervention into automated processes (e.g. \cite{Posada2015}). In addition, combining the strengths of humans and machines in meaningful \ac{HMI} processes targets at increasing manufacturing flexibility in order to reach unprecedented degrees of product customization (e.g. \cite{Zhou2015}). 

Regarding the benefits for the users themselves, \ac{HMI} processes in \ac{I4.0} aim for an improvement of human working conditions and a reduction of professional and, thereby, social exclusion by e.g. compensating for declining cognitive and physical capabilities of older employees, securing employability across different generations of the workforce (e.g. \cite{Saggiomo2016}). Working conditions, in that respect, comprise various factors like occupational health, ergonomics, and safety level at the workplace. Furthermore, regarding employee development and attitude, relying on appropriate and effective means of \ac{HMI} is supposed to foster training efficiency and operators' motivation and engagement at work (e.g. \cite{Stock2016}).

Narrowing the focus to goals for the \acp{UI}, the most important aim for any purposeful \ac{UI}, especially in industrial contexts, is its usability, comprising e.g. the ease of use, intuitiveness in handling, and the resulting user experience and acceptance (e.g. \cite{Gorecky2015}). Besides, \acp{UI} are supposed to deliver transparency of system behavior to the user and shall orchestrate an optimal dynamic task allocation among human and machine entities (e.g. \cite{Trentesaux2016}). Therefore, \ac{UI} development aims at interfaces featuring adaptivity and learning capabilities so that, ultimately, \acp{UI} accurately adapt to the current situational context at run-time (e.g. \cite{Gorecky2014}). Finally, regarding application for purposes of monitoring or providing instructions to operators, an aspired \ac{UI} shall be capable to identify the current process steps performed by workers in order to evaluate their accuracy and intervene in case of deviations from the ideal or planned procedure (e.g. \cite{Longo2017}).

\section{Analysis of the Taxonomy}
\label{chap:results_tax_analysis}

Having answered \ac{RQ1} by defining and describing a comprehensive taxonomy of \ac{HMI} in \ac{I4.0}, the consequential question arises whether there are notable differences in the extent to which the individual elements of the taxonomy are discussed in the articles of the final sample. This question is of particular interest regarding \ac{RQ2} and the identification of focal points and potential research streams in the field. Thereby, such concentrations, if identified, can provide orientation for both researchers and practitioners concerning the state of the art in development and knowledge regarding different aspects and technologies of \ac{HMI} in \ac{I4.0}.

Besides, when analyzing potential research streams on specific sub-aspects of the overall topic reflected in the final sample of this paper, an important question to answer is \ac{RQ3} asking for identifiable patterns in the data collected on sample literature contents. Therefore, the following section provides answers to both \ac{RQ2} and \ac{RQ3} and the respective sub-questions.

\subsection{Results RQ2: Main Foci in Related Research}
\label{sec:rq2}

Based on the insights gained from analyzing literature contents in the final sample, the above-raised question has to be confirmed since there are notable differences in the extent to which different elements of the taxonomy from the same level of hierarchy are covered and discussed among the sample articles. In order to substantiate this result, the following section analyzes the prevalence of top-level to third-level categories of the taxonomy in the final sample.

\begin{figure}[h]
	\centering
	\includegraphics[width=0.9\textwidth]{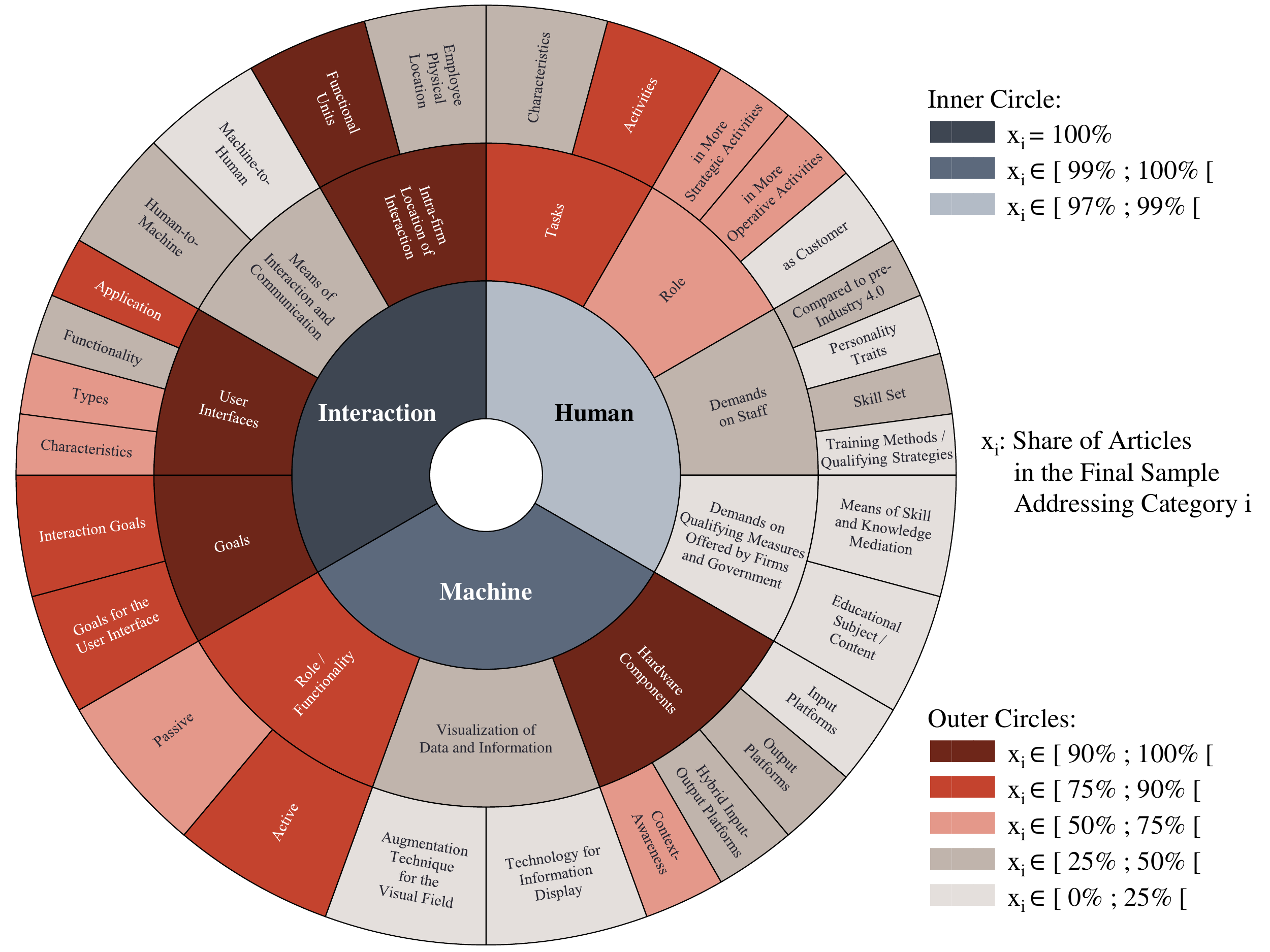}
	\caption{Prevalence of the Individual Categories among the Final Sample}
	\label{figure:taxonomy_frequencies}
\end{figure}

In the form of a heatmap, Figure \ref{figure:taxonomy_frequencies} presents an overview over the frequencies with which the individual categories are addressed in the final sample. To be precise, the color coding of a category indicates the share of articles discussing the respective category by addressing one or more of its lowest-level attributes.

The top-level categories are subject to a special color coding, given that each one is covered in the vast majority of sample articles. Still, it is noteworthy that only the \texttt{Interaction} section is addressed in every paper while 188 of 189 articles discuss the machine entity. Five sample instance, however, cover the topic of \ac{HMI} in \ac{I4.0} without explicitly considering aspects of the taxonomy's \texttt{Human} section. 

Regarding the interim- and bottom-level categories, a five-sector scale has been chosen to visualize differences in the prevalence of individual elements. The lowest three sectors correspond to the first three quartiles while the top quartile is split into a sector covering the top 10\% and the lower 15\%  of the highest quartile. 

What is noticeable at first glance is that there are remarkable differences in the prevalence of individual categories and of larger sections of the taxonomy. In particular, the \texttt{Interaction} section is not just the only top-level category addressed in every sample paper, but it also comprises the majority of the most frequently examined categories. To be precise, seven of overall twelve categories discussed in at least 75\% of sample articles represent a sub-aspect of the interaction part in \ac{I4.0}-related \ac{HMI}.

On the other hand, there are sub-aspects of the taxonomy which seemingly have been studied less frequently and intensively in related scientific literature, meaning that less than half of the sample articles have taken into account attributes of these categories. Regarding the \texttt{Interaction} part, this only concerns the section defining the different means of interaction and communication as well as the separate bottom-level categories of \ac{UI} functionalities and the physical location of interaction. Similarly, in the analysis of the machine entity, the section describing the visualization of data and information and the different types of hardware components except for those ensuring context-awareness have been considered less frequently. In contrast, large parts of the \texttt{Human} section have been studied less often. Those are the demands placed on the qualification measures for employees and those placed on staff members themselves as well as the categories of the human role as a customer and the characteristics of operators' tasks.

More detailed insights regarding the taxonomy elements which have been studied and considered relatively intensively, in contrast, are provided and interpreted in the following subsubsections. Thereby, the results on foci in research related to \ac{HMI} in \ac{I4.0} are supposed to be substantiated in order to answer \ac{RQ2} comprehensively by analyzing and answering each of its three sub-questions. 

\subsubsection{RQ2.1: Focal Points in Research Regarding the Human Aspect}
\label{subsec:rq2_1}

RQ2.1 raises the question which elements of the \texttt{Human} section of the taxonomy are the most prevalent among the scientific articles included in the final sample and thus represent focal points in related research. 

Regarding interim-level categories, it can be inferred from Figure \ref{figure:taxonomy_frequencies} that only the tasks and the role to be assumed by the human operator are discussed in at least half of the sample papers which is why bottom-level categories of such prevalence only exist in these two subsubsections. Those categories are the actual activities and the role of employees in mainly strategic as well as operative activities. 

However, it is worth examining the actual shares of articles addressing the individual categories in more detail. In particular, it seems noteworthy that the demands on staff, with 48.7\%, missed the chosen critical value of 50\% only marginally. Thus, despite the classification on the heatmap, they seem to be regarded as a significant aspect in research when analyzing future \ac{HMI} processes in \ac{I4.0}. 

Nevertheless, the dominance of the human tasks and roles is confirmed as 87.8\% and 71.4\% of sample papers discuss those categories of the \texttt{Human} section, respectively. Especially instances of activities to be performed by humans are quoted regularly (86.2\% of articles) while the strategic and operative roles of humans are discussed in approximately half of the articles, with rather strategic roles being considered slightly more often (53.4\% as opposed to 50.2\%). 

\begin{figure}[h]
	\centering
	\includegraphics[width=13cm]{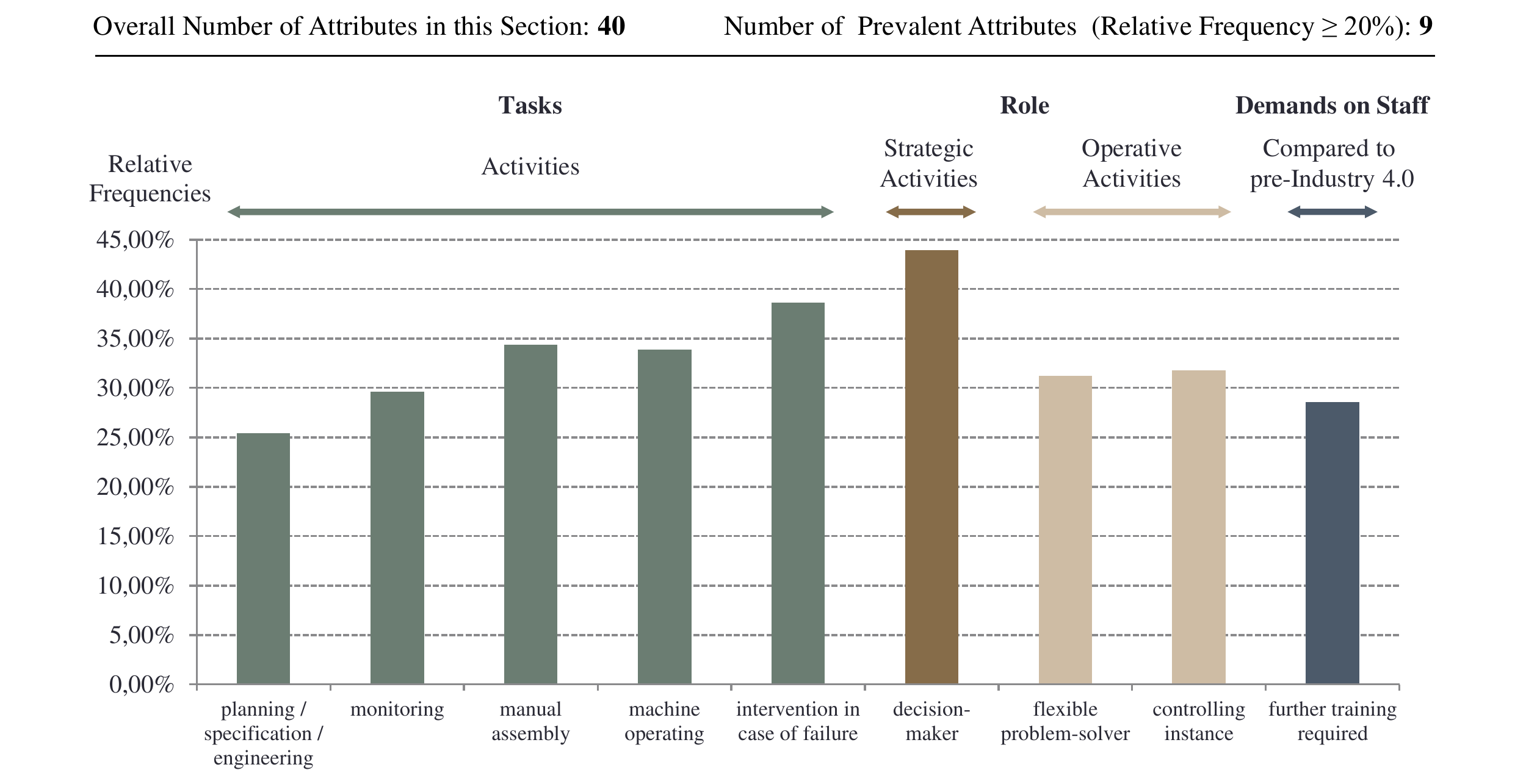}
	\caption{The Most Frequent Attributes of the \texttt{Human} Section}
	\label{figure:frequent_attributes_human}
\end{figure}

Raising the level of detail in the analysis of prevalent taxonomy elements from the \texttt{Human} section, the bottom-level attributes addressed in at least one fifth of sample articles have been identified. The chosen critical value for attributes to be considered as relatively prevalent has been lowered to 20\% to reflect the circumstances of the large number of different attributes and the fact that different authors might address a similar issue by considering different instances from a collection of related attributes. Furthermore, out of a total of 113 attributes of the overall taxonomy, exactly 39 surpass the value of 20\% which means that around one third of attributes are considered as prevalent. This approximately coincides with the evaluation of interim- and bottom-level categories in Section \ref{sec:rq2} where twelve out of 40 categories are addressed in at least 75\% of articles.

As can be inferred from Figure \ref{figure:frequent_attributes_human}, there are, on the one hand, comparatively few attributes from the \texttt{Human} section surpassing the chosen critical value of 20\% while, on the other hand, those attributes which do surpass the 20\% are addressed particularly often. This implies a relatively skewed distribution of prevalence concerning the total of attributes from the \texttt{Human} section. More precisely, while slightly more than a third of all attributes of the taxonomy surpass the value of 20\%, less than a quarter of the attributes from the \texttt{Human} section are discussed so frequently. However, of the nine prevalent attributes from this part of the taxonomy, none is addressed in less than 25\% of the sample papers. 

In general, the impression is reinforced that, regarding a consideration of the human entity, research on \ac{HMI} in \ac{I4.0} is strongly focused on different tasks to be performed by operators. The activity most often quoted is the intervention in case of failures in the production process, corresponding to a large number of studies assigning the role of a flexible problem-solver to human operators. The most prevalent view of the human's role in \ac{I4.0} activities, and at the same time the most frequent attribute of the \texttt{Human} section, however, is that of a strategic decision-maker. Thus, researchers expect employees from various levels of hierarchy to be faced with the challenge and the opportunity to make critical decisions regarding relevant process parameters. Finally, as indicated above, the significance of the demands on staff should not be underrated, considering that 28.6\% of all sample articles expect that transition to \ac{I4.0} will require further training of employees. Besides, more than 19\% of studies mention flexibility as a required personality trait for operators, failing the 20\% value only marginally. 

\subsubsection{RQ2.2: Focal Points in Research Regarding the Machine Aspect}
\label{subsec:rq2_2}

In order to identify the focal points in research regarding the \texttt{Machine} section, i.e. to answer RQ2.2, first the relative frequencies of the corresponding interim- and bottom-level categories are examined. 

While only marginally less than nine in ten sample articles (89.9\%) discuss the role and functionality provided by the machine and 84.7\% and 58.7\% of papers address the \texttt{Active} and \texttt{Passive} sub-categories, respectively, the subsubsection of the hardware components represents a very interesting special case. Although 93.1\% of the articles in the final sample elaborate on the aspect of implemented hardware components, none of their lower-level sub-categories is addressed in at least 75\% of the papers. Hence, the vast majority of articles considers hardware components, however with varying foci. In fact, in total, 80 sample papers (42.3\%) address output and hybrid input-output platforms, respectively, while only 47 papers address both simultaneously. Furthermore, 138 papers (73.0\%) discuss different components ensuring context-awareness of automated systems.

\begin{figure}[h]
	\centering
	\includegraphics[width=13cm]{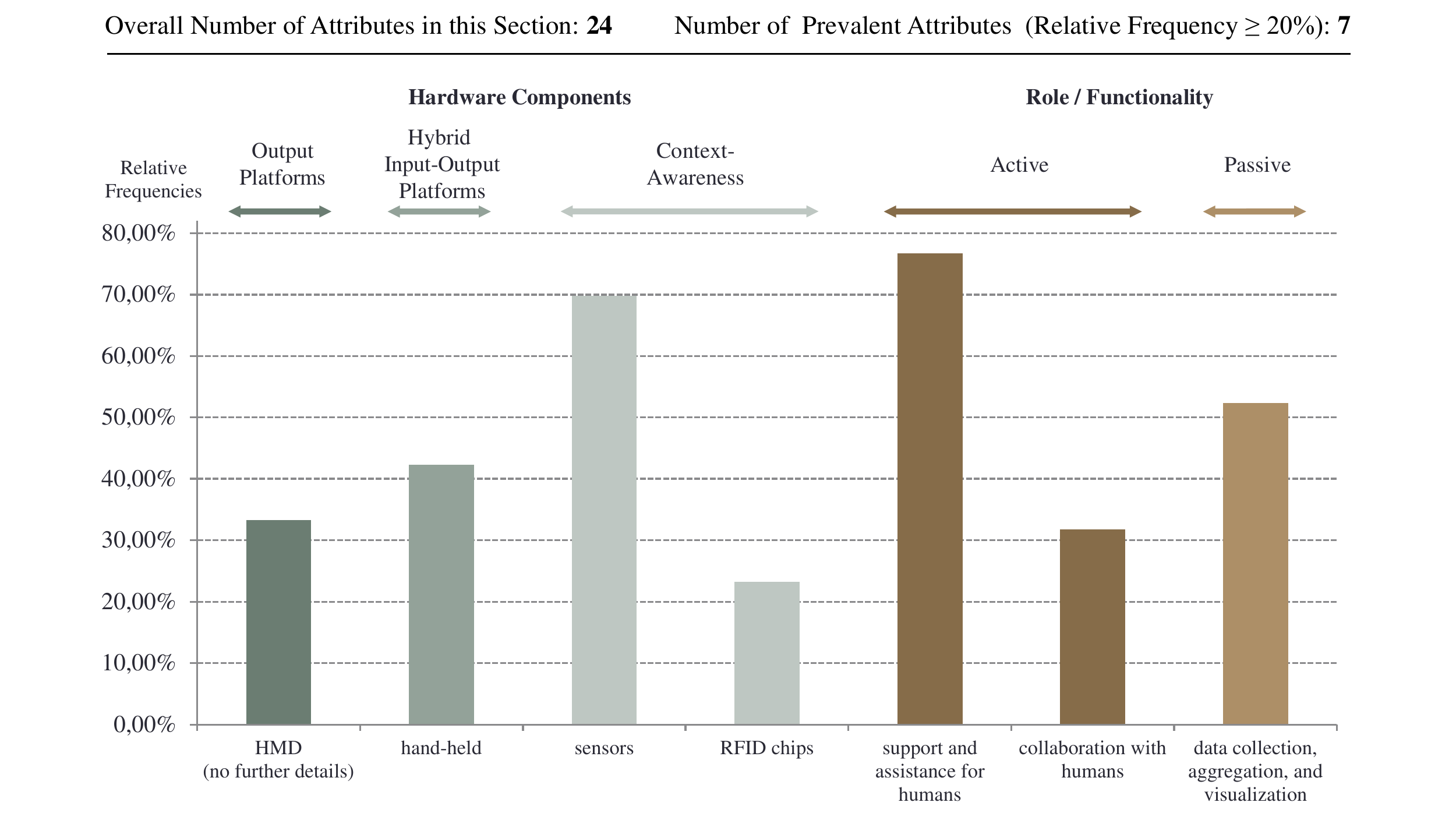}
	\caption{The Most Frequent Attributes of the \texttt{Machine} Section}
	\label{figure:frequent_attributes_machine}
\end{figure}

Examining the prevalence of specific attributes of the \texttt{Machine} section confirms the focal points being output and hybrid input-output platforms, hardware components for context-awareness, as well as active and passive functionalities provided by the machine entity (see Figure \ref{figure:frequent_attributes_machine}). In particular, the machine's role as an active support and assistance system for human operators represents not only the most frequently quoted machine attribute but also the second-most frequent attribute of the overall taxonomy. Besides, the crucial contribution of sensors in achieving context-aware systems and the fundamental role of automated systems of passively collecting, aggregating, and visualizing data and information for human users are also reflected in the data. Finally, the advanced output platform of a not further specified \ac{HMD} is studied intensively in exactly one third of sample publications, partially of an explorative and experimental character, while the use of well-established input-output platforms from the consumer market, i.e. devices like tablet computers and smartphones, for industrial applications has been examined in 42.3\% of all sample articles.

\subsubsection{RQ2.3: Focal Points in Research Regarding the Interaction Aspect}
\label{subsec:rq2_3}

As an answer to the third sub-question of \ac{RQ2}, the most prevalent elements of the \texttt{Interaction} section among the sample articles shall be examined. 

Considering the interim- and bottom-level categories of the taxonomy, it can be stated that more than 90\% of sample articles dedicate their analysis of \ac{HMI} processes in \ac{I4.0} to one or more specific functional units of a manufacturing organization. Still, the sub-aspects of the interaction part most intensively studied are the \acp{UI} deployed and the goals to be achieved, covered by 93.1\% and 95.2\% of articles, respectively. Particularly the interaction goals are addressed in almost 90\% of publications while both the specific goals for the \acp{UI} implemented and their application scenarios are discussed in slightly more than 75\% of sample articles. Besides application, also the other sub-categories of \acp{UI} represent aspects attracting substantial attention in related research. To be precise, the characteristics and types of \acp{UI} implemented are specified in 62.4\% and 65.1\% of sample papers, respectively, while different kinds of functionality provided by those \acp{UI} are discussed in only marginally less than half (49.7\%) of the sample instances.

\begin{figure}[h]
	\centering
	\includegraphics[width=13cm]{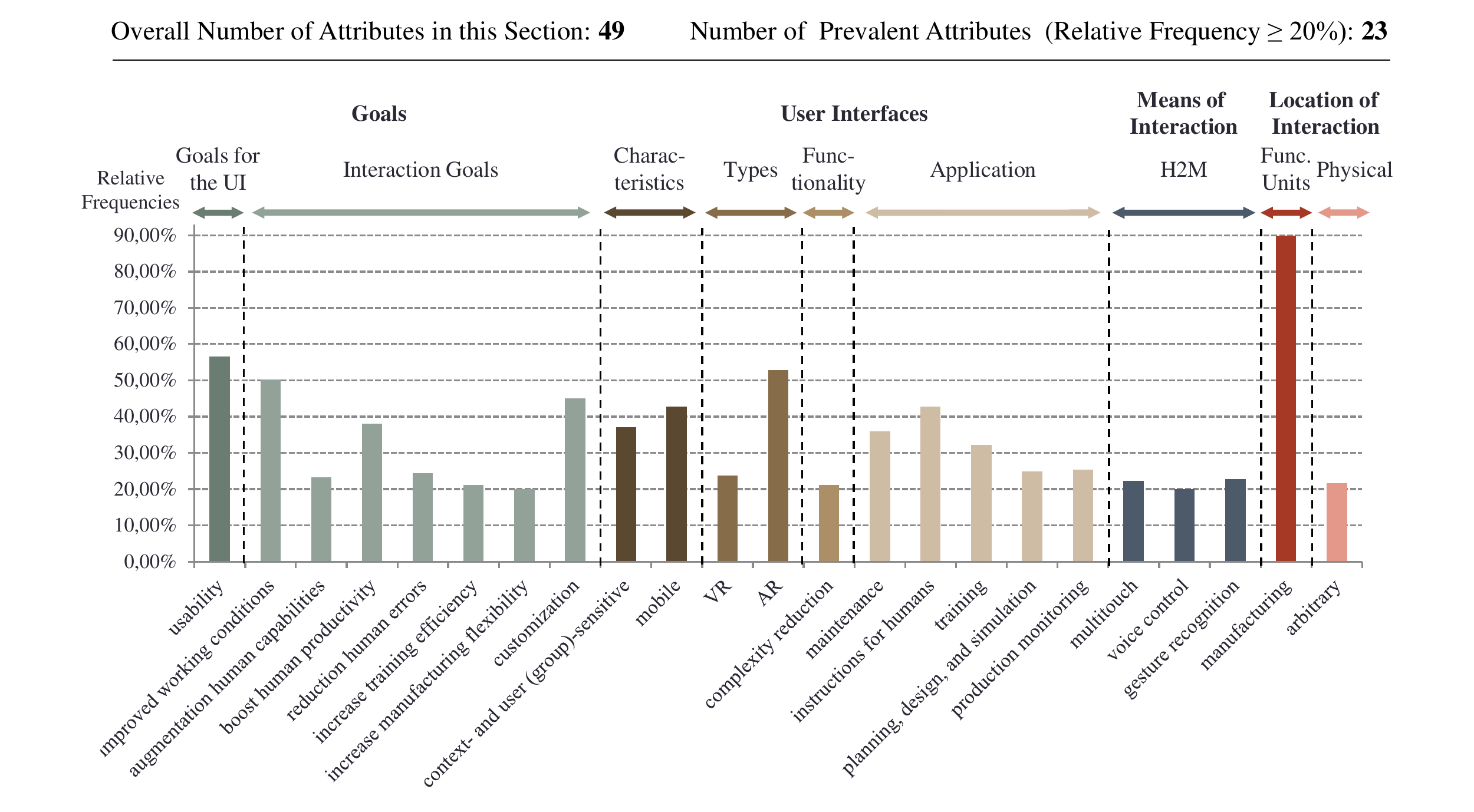}
	\caption{The Most Frequent Attributes of the \texttt{Interaction} Section}
	\label{figure:frequent_attributes_interaction}
\end{figure}

Considering the fact that, overall, approximately one third of all attributes surpass the chosen critial value of 20\% while only 22.5\% of the attributes from the \texttt{Human} section (cf. Figure \ref{figure:frequent_attributes_human}) and 29.2\% of the attributes describing the machine entity (cf. Figure \ref{figure:frequent_attributes_machine}) surpass that value, the share of comparatively prevalent instances among the \texttt{Interaction} section will consequently be higher than a third. This inference is confirmed by Figure \ref{figure:frequent_attributes_interaction}, stating that 46.9\% of attributes from the \texttt{Interaction} section are discussed in more than 20\% of sample papers.

Regarding the much-discussed subsubsection of goals pursued, the instances most often addressed are usability of the \ac{UI} and improved working conditions as well as high degrees of product customization by means of successful \ac{HMI}. Besides, mostly goals concerning the enhancement of human capabilities and performances are frequently addressed, altogether suggesting that a large stream of research on the topic primarily regards \ac{HMI} as a means of efficiently supporting and enhancing the human entity. Concerning \ac{UI} types, it seems striking how many studies are dedicated towards \ac{AR} which represents one of the most often-quoted attributes of the overall taxonomy, reflecting the large number of articles explicitly studying a specific type of \ac{AR} application for industrial purposes. Mainly, those are applications for maintenance purposes or generally visualizing instructions to human operators, explaining their prevalence as separate attributes (36.0\% and 42.9\%, respectively). Apart from \ac{UI} types and applications, their characteristic and functionality of exerting sensitivity and adaptivity towards the situational context seems to be regarded as significant by various authors. Finally, various modern interaction technologies as a means of \ac{H2M} communication are addressed regularly, comprising touch interfaces, natural language interfaces, and gesture recognition, while the vast majority of publications in the final sample studies \ac{HMI} processes occurring within the manufacturing unit of industrial firms.  

\subsection{Results RQ3: Identifiable Patterns in Related Research}
\label{sec:rq3}

After the description of relative frequencies of individual taxonomy elements as an answer to \ac{RQ2} and its sub-questions, the following section presents the results on underlying patterns in the collected data on the coverage of third-level categories in the final sample. Therefore, research questions RQ3.1 and RQ3.2 are analyzed in order to derive association and classification rules from the available set of data, respectively.

\subsubsection{RQ3.1: Derivation of Association Rules from the Content-Related Data}
\label{subsec:rq3_1}

Relating to RQ3.1, data mining techniques are deployed in order to uncover associations among the prevalence of different third-level categories. Thereby, categories can be identified for which the decision of authors on whether or not to address this category in a specific scientific work on \ac{HMI} in \ac{I4.0} exerts a strong association with the authors' respective decision concerning one or more other elements of the taxonomy. Thus, the identification of such relationships, in case they are positive, can be interpreted as a hint that authors from the considered research field regard the associated categories as related contentwise. 

For the purpose of determining association rules with a specified minimum confidence and the highest possible support \cite{Witten2014}, \ac{Weka}'s Apriori algorithm has been applied to the set of data describing the coverage of third-level categories in the sample articles. This means that item sets, i.e. values of specific attributes of the data set or combinations of values of different attributes, are searched which, when observed within a sample instance, appear together with a specific value of another attribute. Minimum confidence has been set to 0.9 which means that for at least 90\% of the sample instances with the specified item set, the value of the associated attribute must equal the specific value in order for this relation to qualify for an association rule. Of all association rules identified, Apriori has been instructed to output the five rules with the highest support, i.e. with the largest number of sample articles for which the identified rule actually holds \cite{Witten2014}.

Unfortunately, applying Apriori to the data set comprising the total of 29 third-level categories as well as the publication year of the respective article as data attributes yields barely meaningful results. Four of the best five association rules identified specify a category whose status (addressed or not addressed) in an article is highly related to the aspect of the human role as a customer not being addressed in the respective article. These associations mainly lack an intuitive rationale and largely follow from the fact that, generally, 93.7\% of sample articles neglect the category of the human role as a customer.

In order to identify a more diverse set of association rules with a more meaningful interpretation, the attribute \texttt{as Customer} has been removed from the data set and the Apriori algorithm has been run a second time in another iteration of the data preparation and modeling phases of \ac{CRISP-DM}. However, the consecutive output based on the reduced data set exerts similar features, mainly naming elements associated with the category \texttt{Functional Units} being addressed in an article. Accordingly, \texttt{Functional Units} has been removed as well before running Apriori again, just like \texttt{Educational Subject/Content} after this third trial.

After removing \texttt{as Customer}, \texttt{Functional Units}, and \texttt{Educational Subject/ Content}, the following five rules have been determined: 
\begin{enumerate}[rightmargin=1.7cm]
	\item{\texttt{Goals for the \ac{UI}} (Addressed) $\Longrightarrow$ \texttt{Interaction Goals} (Addressed)}
	\item{\texttt{Required Skill Set} (Not Addressed) $\Longrightarrow$ \texttt{Required Personality Traits} (Not Addressed)}
	\item{\texttt{Required Skill Set} (Not Addressed) $\Longrightarrow$ \texttt{Means of Skill Mediation} (Not Addressed)}
	\item{\texttt{Augmentation Technique Visual Field} (Not Addressed) $\Longrightarrow$ \texttt{Technology Information Display} (Not Addressed)}
	\item{\texttt{Human Activities\hspace*{0.05cm}/\hspace*{0.05cm}Tasks} (Addressed) $\Longrightarrow$ \texttt{Interaction Goals} (Addressed)}
\end{enumerate}

All of those association rules have an intuitive interpretation for the final deployment phase of the \ac{CRISP-DM} process model. It seems that authors discussing the goals for applied \acp{UI} mostly regard them in a broader perspective, considering how they contribute to the achievement of general goals for \ac{HMI} in \ac{I4.0}. These interaction goals also seem to be of high relevance when discussing future human tasks, consistent with the notion that augmented operators will be supported by technology in performing their tasks (cf. \cite{Weyer2015}). On the other hand, not considering the required human skill set obviously makes a discussion of means of mediating skills to employees redundant while it also seems to imply that the respective authors forego a discussion of required human characteristics in general, both regarding skill set and personality traits. Finally, a publication not discussing augmentation techniques for the visual field seems to indicate that the authors generally concentrate on sub-aspects other than the visualization of data and information to human operators.

\subsubsection{RQ3.2: Derivation of Classification Rules from the Content-Related Data}
\label{subsec:rq3_2}

As a final step in analyzing the data collected on the sample literature, \ac{Weka}'s JRip classifier is supposed to identify potential classification rules regarding different third-level categories of the taxonomy. Such classifications can help researchers determine whether related research would generally be likely to consider a specific aspect of the topic given the remaining aspects supposed to be addressed in the study. Based on this information, unintended disregarding of important sub-aspects in the paper can be avoided or authors can deliberately concatenate different aspects of the overall topic not widely studied in combination yet.

Compared to other variants of classification, JRip regularly delivers a very lean and less complex model in the form of a manageable set of consecutive rules classifying an instance based on the values of a specific set of its attributes \cite{Witten2014}. Particularly with regard to the number of third-level categories of the taxonomy (29), models requiring the values of all attributes except for the class as input to predict an instance's class value as well as tree classifiers with too many branches are impractical and impede an intuitive interpretation of the classification model.

Overall, for thirteen out of 30 attributes in the data set, the classification model produced by the JRip algorithm is discarded because a simplistic and unconditional classification based on the rule delivered by ZeroR implies a higher estimated accuracy. More precisely, the model of rules produced by JRip is not considered meaningful or as adding valuable information if its accuracy is estimated to be lower than that of an unconditional classification rule relying solely on the most prevalent class value in the training set \cite{Witten2013a}.

Furthermore, the identified classification rules for another six class attributes are not considered for further analysis (i.e. deployment) because their estimated classification accuracy does not exceed that of the corresponding rule determined by ZeroR by more than 3.6 percentage points while for the remaining eleven classification models, this difference is larger than five percentage points. Thus, for those eleven class attributes, the chances are higher that JRip provides a significantly higher mean classification accuracy than ZeroR.

As stated in Section \ref{subsec:analysis_data_mining}, \ac{ROC} area values of the accepted and considered models are examined as another means of verifying their meaningfulness. For a random and thus uninformed or worthless classifier, \ac{ROC} area would be expected to be close to 0.5 while a perfect classifier would exhibit a \ac{ROC} area of 1 given that it would always predict articles actually addressing a specific category to address the category (true positive rate of 1) and would never predict articles not discussing a specific category to address this element (false positive rate of 0) \cite{Powers2011}. 

Considering that the unconditional classification models delivered by ZeroR predict all instances to have the same class value, true and false positive rate for a specific class value are equal (either 0 or 1) which is also expected for a random classifier (rates are expected to be equal, not necessarily 0 or 1) \cite{Powers2011}. Thus, these uninformed models are expected to exhibit a \ac{ROC} area value similar to a random classifier. A classification model based on JRip, on the other hand, if meaningful and informed, should be closer to the perfect classifier's true and false positive rates and exhibit a \ac{ROC} area greater than 0.5 (cf.~\cite{Powers2011}).

Indeed, examination of \ac{ROC} area values for the eleven sets of rules considered for analysis reveals that those classification models exhibit a respective value of at least 0.61, considerably larger than the expected baseline case of approximately 0.5 for a random classifier \cite{Powers2011}. Thus, those identified classification models are considered to add value by providing some informed knowledge about the classification of research articles related to the topic of \ac{HMI} in \ac{I4.0}.

The five most intuitive and meaningful sets of classification rules among the eleven considered models, regarding interpretability of the rules, classify research articles regarding one of the following third-level categories, respectively: the characteristics of human tasks, the demands on staff compared to pre-\ac{I4.0}, hardware output platforms, \ac{UI} functionality, or means of \ac{H2M} communication. Table \ref{table_classification_rules} presents the list of classification rules for these class attributes, also including the sets of rules for the other six accepted and considered classification models. 
%Appendix \ref{sec:apx_classification_rules} lists the six remaining accepted classification models which have not been considered for further analysis.

\begin{table}[h]
	%\begin{addmargin}{\dimexpr -\oddsidemargin-0.4in\relax}
	\resizebox{\textwidth}{!}{
		\begin{tabular}{p{1.8cm} p{0.5cm} p{0.5cm} C{0.4cm} C{0.4cm} p{11cm}}
			\textbf{\scriptsize{Class}} & \multicolumn{2}{c}{\textbf{\scriptsize{Accuracy}}} & \multicolumn{2}{c}{\textbf{\scriptsize{\acs{ROC} Area}}} & \textbf{\scriptsize{Rules (JRip)}} \\
			&	\tiny{ZeroR} &	\textbf{\tiny{JRip}} &	\tiny{ZeroR} &	\textbf{\tiny{JRip}} &	 \tiny{(A. $=$ Addressed; N.A. $=$ Not Addressed)} \\ \hline \hline
			\textbf{\tiny{Task}} &	\tiny{69.8\%} &	\tiny{75.4\%} & \tiny{0.48} & \tiny{0.69} &	\textbf{\tiny{Demands Compared to pre-I4.0 (A.) \& Operative Role (A.) $\Longrightarrow$ Task Character. (A.)}} \\
			\textbf{\tiny{Characteristics}} &	\tiny{} &	\tiny{} &	\tiny{} &	\tiny{} &	\textbf{\tiny{ $\Longrightarrow$ Task Characteristics (N.A.)}} \\ \hline
			\tiny{Strategic Role} &	\tiny{53.4\%} &	\tiny{61.5\%} &	\tiny{0.49} &	\tiny{0.62} &	\tiny{Technology Information Display (A.) \& Operative Role (N.A.) $\Longrightarrow$ Strategic Role (N.A.)} \\
			\tiny{} &	\tiny{} &	\tiny{} &	\tiny{} &	\tiny{} &	\tiny{Input Platforms (A.) \& UI Characteristics (N.A.) $\Longrightarrow$ Strategic Role (N.A.)} \\
			\tiny{} &	\tiny{} &	\tiny{} &	\tiny{} &	\tiny{} &	\tiny{Operative Role (N.A.) \& Context-Awareness (N.A.) \& Human Activities/Tasks (A.) \& M2H Communication (N.A.) $\Longrightarrow$ Strategic Role (N.A.)} \\
			\tiny{} &	\tiny{} &	\tiny{} &	\tiny{} &	\tiny{} &	\tiny{M2H Communication (A.) \& Demands Compared to pre-I4.0 (N.A.) \& Hybrid Platforms (N.A.) \& Active (A.) $\Longrightarrow$ Strategic Role (N.A.)} \\
			\tiny{} &	\tiny{} &	\tiny{} &	\tiny{} &	\tiny{} &	\tiny{ $\Longrightarrow$ Strategic Role (A.)} \\ \hline
			\tiny{Operative Role} &	\tiny{50.3\%} &	\tiny{65.4\%} &	\tiny{0.47} &	\tiny{0.66} &	\tiny{Task Characteristics (N.A.) \& Required Personality Traits (N.A.) $\Longrightarrow$ Operative Role (N.A.)} \\
			\tiny{} &	\tiny{} &	\tiny{} &	\tiny{} &	\tiny{} &	\tiny{Required Skill Set (N.A.) \& UI Types (A.) \& Output Platforms (N.A.) $\Longrightarrow$ Operative Role (N.A.)} \\
			\tiny{} &	\tiny{} &	\tiny{} &	\tiny{} &	\tiny{} &	\tiny{as Customer (A.) \& year $=$ 2016 $\Longrightarrow$ Operative Role (N.A.)} \\
			\tiny{} &	\tiny{} &	\tiny{} &	\tiny{} &	\tiny{} &	\tiny{Augm. Techn. Vis. Field (A.) \& Active (A.) \& Req. Person. Traits (A.) $\Longrightarrow$ Oper. Role (N.A.)} \\
			\tiny{} &	\tiny{} &	\tiny{} &	\tiny{} &	\tiny{} &	\tiny{ $\Longrightarrow$ Operative Role (A.)} \\ \hline
			\textbf{\tiny{Demands}} &	\tiny{64.0\%} &	\tiny{81.5\%} &	\tiny{0.49} &	\tiny{0.80} &	\textbf{\tiny{Required Skill Set (A.) $\Longrightarrow$ Demands Compared to pre-I4.0 (A.)}} \\
			\textbf{\tiny{Compared}} &	\tiny{} &	\tiny{} &	\tiny{} &	\tiny{} &	\textbf{\tiny{Educational Subject/Content (A.) $\Longrightarrow$ Demands Compared to pre-I4.0 (A.)}} \\
			\textbf{\tiny{to pre-I4.0}} &	\tiny{} &	\tiny{} &	\tiny{} &	\tiny{} &	\textbf{\tiny{ $\Longrightarrow$ Demands Compared to pre-I4.0 (N.A.)}} \\ \hline
			\textbf{\tiny{Output}} &	\tiny{57.7\%} &	\tiny{76.9\%} &	\tiny{0.50} &	\tiny{0.79} &	\textbf{\tiny{UI Types (A.) \& Augmentation Technique Visual Field (A.) $\Longrightarrow$ Output Platforms (A.)}} \\
			\textbf{\tiny{Platforms}} &	\tiny{} &	\tiny{} &	\tiny{} &	\tiny{} &	\textbf{\tiny{UI Types (A.) \& Operative Role (A.) $\Longrightarrow$ Output Platforms (A.)}} \\
			\tiny{} &	\tiny{} &	\tiny{} &	\tiny{} &	\tiny{} &	\textbf{\tiny{ $\Longrightarrow$ Output Platforms (N.A.)}} \\ \hline
			\tiny{Hybrid} &	\tiny{57.7\%} &	\tiny{69.0\%} &	\tiny{0.50} &	\tiny{0.70} &	\tiny{UI Characteristics (A.) \& Output Platforms (A.) $\Longrightarrow$ Hybrid Platforms (A.)} \\
			\tiny{Platforms} &	\tiny{} &	\tiny{} &	\tiny{} &	\tiny{} &	\tiny{UI Characteristics (A.) \& Functional Units (A.) \& Input Platforms (N.A.) \& Required Personality Traits (N.A.) $\Longrightarrow$ Hybrid Platforms (A.)} \\
			\tiny{} &	\tiny{} &	\tiny{} &	\tiny{} &	\tiny{} &	\tiny{ $\Longrightarrow$ Hybrid Platforms (N.A.)} \\ \hline
			\tiny{UI} &	\tiny{62.4\%} &	\tiny{67.8\%} &	\tiny{0.49} &	\tiny{0.69} &	\tiny{Hybrid Platforms (N.A.) \& Passive Functionality (N.A.) $\Longrightarrow$ UI Characteristics (N.A.)} \\
			\tiny{Characteristics} &	\tiny{} &	\tiny{} &	\tiny{} &	\tiny{} &	\tiny{Hybrid Platforms (N.A.) \& as Customer (A.) $\Longrightarrow$ UI Characteristics (N.A.)} \\
			\tiny{} &	\tiny{} &	\tiny{} &	\tiny{} &	\tiny{} &	\tiny{ $\Longrightarrow$ UI Characteristics (A.)} \\ \hline
			\tiny{UI Types} &	\tiny{65.1\%} &	\tiny{84.2\%} &	\tiny{0.47} &	\tiny{0.83} &	\tiny{Output Platforms (N.A.) \& UI Application (N.A.) $\Longrightarrow$ UI Types (N.A.)} \\
			\tiny{} &	\tiny{} &	\tiny{} &	\tiny{} &	\tiny{} &	\tiny{Output Platforms (N.A.) \& Operative Role (A.) \& Physical Location (N.A.) $\Longrightarrow$ UI Types (N.A.)} \\
			\tiny{} &	\tiny{} &	\tiny{} &	\tiny{} &	\tiny{} &	\tiny{Goals for the UI (N.A.) \& Physical Location (A.) $\Longrightarrow$ UI Types (N.A.)} \\
			\tiny{} &	\tiny{} &	\tiny{} &	\tiny{} &	\tiny{} &	\tiny{ $\Longrightarrow$ UI Types (A.)} \\ \hline
			\textbf{\tiny{UI}} &	\tiny{47.6\%} &	\tiny{62.5\%} &	\tiny{0.47} &	\tiny{0.61} &	\textbf{\tiny{H2M Communication (A.) \& M2H Communication (A.) $\Longrightarrow$ UI Functionality (A.)}} \\
			\textbf{\tiny{Functionality}} &	\tiny{} &	\tiny{} &	\tiny{} &	\tiny{} &	\textbf{\tiny{Hybrid Platforms (A.) $\Longrightarrow$ UI Functionality (A.)}} \\
			\tiny{} &	\tiny{} &	\tiny{} &	\tiny{} &	\tiny{} &	\textbf{\tiny{ $\Longrightarrow$ UI Functionality (N.A.)}} \\ \hline
			\textbf{\tiny{H2M}} &	\tiny{61.4\%} &	\tiny{69.2\%} &	\tiny{0.48} &	\tiny{0.65} &	\textbf{\tiny{Hybrid Platforms (A.) \& Operative Role (A.) $\Longrightarrow$ H2M Communication (A.)}} \\
			\textbf{\tiny{Communication}} &	\tiny{} &	\tiny{} &	\tiny{} &	\tiny{} &	\textbf{\tiny{Input Platforms (A.) \& Output Platforms (N.A.) $\Longrightarrow$ H2M Communication (A.)}} \\
			\tiny{} &	\tiny{} &	\tiny{} &	\tiny{} &	\tiny{} &	\textbf{\tiny{UI Types (A.) \& UI Functionality (A.) \& M2H Comm. (A.) $\Longrightarrow$ H2M Comm. (A.)}} \\
			\tiny{} &	\tiny{} &	\tiny{} &	\tiny{} &	\tiny{} &	\textbf{\tiny{ $\Longrightarrow$ H2M Communication (N.A.)}} \\ \hline
			\tiny{Physical} &	\tiny{69.8\%} &	\tiny{76.1\%} &	\tiny{0.48} &	\tiny{0.63} &	\tiny{Input Platforms (A.) \& Passive Functionality (N.A.) $\Longrightarrow$ Physical Location (A.)} \\
			\tiny{Location} &	\tiny{} &	\tiny{} &	\tiny{} &	\tiny{} &	\tiny{ $\Longrightarrow$ Physical Location (N.A.)} \\ \hline
		\end{tabular}}
	%\end{addmargin}
	\caption{Accepted and Considered Classification Models Produced by JRip}
	\label{table_classification_rules}
\end{table}

\textbf{Characteristics of Human Tasks.} The model for characteristics of human tasks (cf. Table \ref{table_classification_rules}) is interesting as it suggests that authors regard the changes in demands on staff brought by \ac{I4.0} for employees with a more operative role in the operations to be linked to the characteristics of their actual tasks under \ac{I4.0}.

\textbf{Demands on Staff Compared to pre-\ac{I4.0}.} In relation to those changes in demands on staff, in turn, when discussing required skills of employees, authors seem to imply a change compared to pre-\ac{I4.0}, causing them to compare demands before and after inception of \ac{I4.0} (cf. Table \ref{table_classification_rules}).

\textbf{Output Platforms.} The classification rules for coverage of output platforms (cf. Table \ref{table_classification_rules}) seem intuitively sensible as different types of \acp{UI} can be seen as commonly related to differing output platforms, e.g. when regarding \ac{VR} as rather linked to \acp{HMD} than to stationary screens while the opposite applies to traditional \acp{GUI}. Therefore, a discussion of output platforms due to the consideration of specific \acp{UI} appears plausible. The additional condition of addressing the augmentation technique for the visual field is suitable in the sense that it hints towards the fact that the study reflects on the technological aspects of visualizing information in its analysis of \ac{UI} types, making the consideration of output platforms consequent. Varying human operative roles, on the other hand, might require the consideration of different suitable types of output platforms.

\textbf{\ac{UI} Functionality.} The model for \ac{UI} functionality has the lowest \ac{ROC} area value and degree of estimated accuracy (62.5\%) among the five models discussed (cf. Table \ref{table_classification_rules}). Still, a discussion of the functionalities provided by the implemented \acp{UI} seems consequent if possible means of \ac{HMI} in both directions are examined or when hybrid platforms are considered which typically provide a broad range of \ac{UI} functionalities.

\textbf{Means of \ac{H2M} Communication.} The first rule of the model for \ac{H2M} communication (cf. Table \ref{table_classification_rules}) is plausible given that hybrid platforms regularly offer multiple means of \ac{H2M} communication while depending on the specific operative role, different forms of interaction from the operator to the automated system might be more appropriate. The second rule is perfectly plausible as well, considering that a discussion of input platforms while neglecting corresponding output platforms suggests a focus on \ac{H2M} communication. Finally, regarding the third rule, a concurrent analysis of types and functionalities of \acp{UI} as well as \ac{M2H} communication reflects a broad examination of \ac{HMI} means and tools, potentially requiring also a consideration of \ac{H2M} communication for completion.

\section{Conclusion}
\label{chap:conclusion}

Having presented the applied methodology, the findings which resulted from said procedure, as well as the implications and limitations of these results, the last chapter of this thesis is supposed to provide a summarizing overview of the most important insights, together with concluding remarks. Building upon that, consequent and open research issues are identified which represent opportunities for further research on the topic of \ac{HMI} in \ac{I4.0}.

\subsection{Summary}
\label{sec:summary}

The aim of this thesis has been to analyze a comprehensive set of scientific literature relevant to the topic of \ac{HMI} in \ac{I4.0} and to develop, based upon this foundation, a taxonomy representing the current state of the art regarding research in the area of the topic. Furthermore, the insights on the taxonomy and data collected on its coverage in related literature should serve as a basis to identify focal points and patterns in research on \ac{HMI} in \ac{I4.0}.

The taxonomy developed as an answer to \ac{RQ1} structures the topic in a four-layer hierarchy consisting of three levels of categories and a bottom-level collection of attributes representing instances or specifications of the third-level categories. On the highest rank, the topic is clustered into three sections representing the human actor, the machine entity, and aspects defining the process of interaction between both. On the second and third level, those three main categories are defined in more detail in terms of their constituting aspects.

In order to capture the human aspect in \ac{I4.0}-related \ac{HMI}, operators' tasks as well as the human role in \ac{I4.0} operations are defined. Moreover, the taxonomy determines demands on employees and the qualifying measures to be offered by firms and government in order to prepare an appropriate \ac{I4.0} workforce as the final sections of the \texttt{Human} part.

In describing the machine entity, the taxonomy considers passive and active roles and functionalities assumed or offered by the automated system as well as the available techniques and technologies representing the machine's means of visualizing data and information to human users. Apart from that, the taxonomy includes an analysis of the available hardware components needed to provide such functionality and visualization services.

The third main section of the taxonomy defines the interaction processes among human and machine agents under \ac{I4.0} in terms of different types of goals to be achieved in the interaction, the \acp{UI} deployed, the means of interaction and communication, as well as the location of interaction.

The analysis of the sample literature's coverage of individual taxonomy elements in order to answer \ac{RQ2} has revealed that the interim-level categories of \acp{UI} and goals in the interaction as well as the implemented hardware components represent focal points in \ac{HMI}-related research in the \ac{I4.0} area. Only slightly less prevalent, discussions of the human tasks and the functionality provided by the machine entity also seem to be regarded as highly significant in studying \ac{HMI} in \ac{I4.0} by the majority of authors in the final sample. Thus, a main stream in related research seems to approach the topic by distinguishing tasks to be performed by humans from the role and functionality provided by automated systems and enabled by various hardware components. Based on that, the integration of those two types of entities in an interaction process is supposed to serve specific purposes and goals, facilitated by the implementation of appropriate \acp{UI}. In general, the vast majority of sample articles conducts such analyses and discussions of \ac{HMI} in \ac{I4.0} with respect to a context of manufacturing activities, revealing a focus of researchers on the industrial core activity of production.

In the course of answering \ac{RQ3}, application of data mining techniques on the data reflecting the sample papers' coverage of third-level categories reveals a set of association rules represented in the data. These rules describe pairs or sets of third-level categories exhibiting interrelations in the fact whether or not they are addressed in individual sample articles. Finally, using \ac{Weka}'s JRip algorithm, seventeen sets of classification rules have been identified serving as a potential support for researchers in deciding whether they should include a discussion of a specific aspect in a study dedicated towards \ac{HMI} in \ac{I4.0}. The most intuitive sets of rules classify research articles with regard to the characteristics of human tasks, the demands on staff compared to pre-\ac{I4.0}, output platforms, \ac{UI} functionality, and means of \ac{H2M} communication.

The potential implications of these results extend to future employees and firms in the industrial sector, facing changes in the working environment in general and challenges regarding measures and investments to be taken in preparing for the fourth industrial revolution, respectively, as well as to governments and customers of industrial firms. However, the implications derived from the study's results need to be interpreted with care by taking into account the restricted scope of the study with respect to the number of publications related to the topic which could be included in the initial sample and the limitations of the applied methodology.

Altogether, the findings in the course of this thesis reflect the complexity of the overall topic of \ac{HMI} in \ac{I4.0} by organizing it in a four-level taxonomy of categories and attributes. Still, the taxonomy, and thus the topic, exhibits a clear structure on different levels of detail whose individual elements and related sections are reflected in related research to varying extents. Examining the sub-aspects of the taxonomy studied less frequently, judging based on the final sample of relevant literature, reveals open research issues representing opportunities for future research which the following final section briefly outlines.

\subsection{Opportunities for Future Research}
\label{sec:future_research}

When examining the machine section of the taxonomy more closely, it is noticeable that a substantial amount of sample articles, exactly one third to be precise, incorporates the application of \acp{HMD} of a not further specified type as the most prevalent attribute of the output platforms category. At the same time, none of the attributes or even the third-level categories themselves of input platforms, technology for information display, and augmentation technique for the visual field is addressed in more than 22.3\% of articles. Considered in combination, this suggests that there might be room left for analyzing the most appropriate input platforms to be employed in conjunction with \acp{HMD} as the corresponding output platform for \ac{I4.0}-related \ac{HMI} processes. Likewise, an examination of the most adequate techniques and technologies for information visualization implemented on an \ac{HMD} output platform might yield interesting and useful insights.

Regarding the much-discussed aspect of \acp{UI}, an apparent phenomenon among this thesis' final sample is the large number of articles studying a specific industrial \ac{AR} application, respectively. Nonetheless, the insights on longer-term effects and applicability of \ac{AR} for such scenarios are rare. Therefore, in light of the apparent significance of \ac{AR} applications in related research, conducting a long-term experimental study on an industrial \ac{AR} application might represent an important opportunity for future research.

Besides, despite the central role of human operators in \ac{I4.0} activities and 28.6\% of sample articles stressing the need for further training of employees to enable successful implementation of the related processes, none of the third-level categories concerned with aspects of staff training and qualification is addressed in at least 25\% sample papers. Therefore, further studies on appropriate measures of education and qualification of future and current members of the industrial workforce might be of significant value to public and corporate decision-makers.

Finally, the analyses and findings in the course of this thesis might be further advanced by using technical methods like machine learning algorithms to cluster the bottom-level attributes based on observed sample data. This way, categories and a structure for a potential alternative design of a taxonomy even better reflected in related scientific contributions might be identified. Furthermore, the implemented analyses of the collected data might be intensified by conducting more iterations of the \ac{CRISP-DM} process model, searching for classification models better suited for the available set of data.

%%
%% The acknowledgments section is defined using the "acks" environment
%% (and NOT an unnumbered section). This ensures the proper
%% identification of the section in the article metadata, and the
%% consistent spelling of the heading.
%\begin{acks}
%To Robert, for the bagels and explaining CMYK and color spaces.
%\end{acks}

%%
%% The next two lines define the bibliography style to be used, and
%% the bibliography file.
\bibliographystyle{ACM-Reference-Format}
\bibliography{hmi_industry40_survey}

%%
%% If your work has an appendix, this is the place to put it.
\begin{comment}
\appendix

\section{Research Methods}

\subsection{Part One}

Lorem ipsum dolor sit amet, consectetur adipiscing elit. Morbi
malesuada, quam in pulvinar varius, metus nunc fermentum urna, id
sollicitudin purus odio sit amet enim. Aliquam ullamcorper eu ipsum
vel mollis. Curabitur quis dictum nisl. Phasellus vel semper risus, et
lacinia dolor. Integer ultricies commodo sem nec semper.

\subsection{Part Two}

Etiam commodo feugiat nisl pulvinar pellentesque. Etiam auctor sodales
ligula, non varius nibh pulvinar semper. Suspendisse nec lectus non
ipsum convallis congue hendrerit vitae sapien. Donec at laoreet
eros. Vivamus non purus placerat, scelerisque diam eu, cursus
ante. Etiam aliquam tortor auctor efficitur mattis.

\section{Online Resources}

Nam id fermentum dui. Suspendisse sagittis tortor a nulla mollis, in
pulvinar ex pretium. Sed interdum orci quis metus euismod, et sagittis
enim maximus. Vestibulum gravida massa ut felis suscipit
congue. Quisque mattis elit a risus ultrices commodo venenatis eget
dui. Etiam sagittis eleifend elementum.

Nam interdum magna at lectus dignissim, ac dignissim lorem
rhoncus. Maecenas eu arcu ac neque placerat aliquam. Nunc pulvinar
massa et mattis lacinia.
\end{comment}

\end{document}

%% file: acronyms.tex
\acrodef{3D}{three-dimensional}
\acrodef{AR}{Augmented Reality}
\acrodef{CPPS}{Cyber-Physical Production System}
\acrodef{CPS}{Cyber-Physical System}
\acrodef{CRISP-DM}{CRoss Industry Standard Process for Data Mining}
\acrodef{GPS}{Global Positioning System}
\acrodef{GUI}{graphical user interface}
\acrodef{H2M}{human-to-machine}
\acrodef{HMD}{head-mounted display}
\acrodef{HMI}{human-machine interaction}
\acrodef{IEEE}{Institute of Electrical and Electronics Engineers}
\acrodef{I4.0}{Industry 4.0}
\acrodef{IoT}{Internet of Things}
\acrodef{M2H}{machine-to-human}
\acrodef{MINT}{mathematics, informatics, natural sciences, and technology}
\acrodef{NUI}{natural user interface}
\acrodef{RFID}{radio-frequency identification}
\acrodef{RQ1}{Research Question 1}
\acrodef{RQ2}{Research Question 2}
\acrodef{RQ3}{Research Question 3}
\acrodef{ROC}{Receiver Operating Characteristics}
\acrodef{UI}{user interface}
\acrodef{VR}{Virtual Reality}
\acrodef{Weka}{Waikato Environment for Knowledge Analysis}
\acrodef{WIMP}{windows, icons, menus, pointer}